\documentclass[12pt]{article}
\usepackage{amsmath}
\usepackage{epsfig}
\usepackage{amsfonts}
\usepackage{amssymb}
\usepackage{environ}

\def\beq{\begin{equation}}
\def\eeq{\end{equation}}
\def\bea{\begin{eqnarray}}
\def\eea{\end{eqnarray}}
\def\beas{\begin{eqnarray*}}
\def\eeas{\end{eqnarray*}}
\def\Tr{{\rm Tr}}

\def\Xint#1{\mathchoice
   {\XXint\displaystyle\textstyle{#1}}%
   {\XXint\textstyle\scriptstyle{#1}}%
   {\XXint\scriptstyle\scriptscriptstyle{#1}}%
   {\XXint\scriptscriptstyle\scriptscriptstyle{#1}}%
   \!\int}
\def\XXint#1#2#3{{\setbox0=\hbox{$#1{#2#3}{\int}$}
     \vcenter{\hbox{$#2#3$}}\kern-.5\wd0}}

\def\dashint{\Xint-}

\def\sos{{\rm sos}}
\def\cin{{\rm cin}}

\begin{document}


\begin{titlepage}

\begin{centering}

\vspace*{3cm}

{\Large\bf Solving 
           two-dimensional adjoint QCD with a basis-function approach}

\vspace*{1.5cm}

{\bf Uwe Trittmann}
\vspace*{0.5cm}

{\sl Department of Physics\\
Otterbein University\\
Westerville, OH 43081, USA}

\vspace*{1cm}

\today

\vspace*{2cm}

{\large Abstract}

\vspace*{1cm}

\end{centering}

We apply a method (``eLCQ'') to find the asymptotic
spectrum of a Hamiltonian from its symmetries to two-dimensional adjoint
QCD.
Streamlining the approach, we construct a complete
set of asymptotic eigenfunctions in all parton sectors and
use it in a basis-function approach to find
the spectrum of the full theory. We are able to reproduce
previous results including
the degeneracy of fermionic and bosonic masses at the supersymmetric
point, and to understand the properties of the lowest states in the
massless theory.  

The approach taken here is continuous at fixed parton number, and
therefore complementary to standard (DLCQ-like) formulations.
Despite its limitation to rather small parton numbers, it
can be used to test and validate conclusions of other frameworks
in an independent way.

\vspace{0.5cm}

\end{titlepage}
\newpage


\section{Introduction}
\label{SecIntro}

In the last years there has been renewed interest in two-dimensional
adjoint quantum chromodynamics (QCD$_{2A}$), as new methods and
computational resources have become available
\cite{Katz,Jacobson,Kleb21,Kleb22}.  The theory has been generalized
and thereby become interesting for other subfields \cite{GHKSS}.  It
is simpler in the large $N$ limit, but forays into finite $N$
calculation \cite{Anton} have recently been put on a solid foundation
\cite{Kleb22}.  The theory of {\em massive} adjoint fermions is
confining and possesses a supersymmetric point at $m_{adj}=g^2N/\pi$
\cite{Kutasov94}. Its features are well reproduced in numerical
studies starting with \cite{DalleyKlebanov, BDK}. The {\em massless}
theory can be bosonized, i.e.~cast into a theory of currents obeying a
Kac-Moody algebra \cite{Witten84}. For several reasons it is the more
interesting and challenging version
of QCD$_{2A}$. For instance, pair production is
likely important, and thus sectors of different parton\footnote{ A
  {\em parton} is a current or a fermion, depending on whether the
  theory is bosonized or not.} number are coupled. Lately, there have
been discussions \cite{Jacobson,Smilga21,Kleb21} about the
conjectured screening-confinement transition \cite{GrossKlebMatSmilga}
as the fermion mass
vanishes, $m_{adj}\rightarrow 0$.  To extract the true single-particle
content of the theory has been difficult because single-particle states
cannot be identified with single-trace states \cite{KutasovSchwimmer}.
As was shown numerically starting with \cite{GHK}, at least some of
the single-trace eigenstates of QCD$_2$ must be categorized as
multi-particle states. This was corroborated in \cite{UT1} by the fact
that in the bosonized theory exact multi-particle states, i.e.~states
with the same mass as a state of non-interacting partons, are
projected out.

To see how the present approach can better our understanding
of QCD$_{2A}$, it is helpful to review some of the
literature on the subject in more detail. 
For starters, QCD$_{2A}$ is a 1+1 dimensional SU($N$) gauge theory coupled
to a (Majorana) fermion in the adjoint representation. 
It is a modification of the 't Hooft
model \cite{tHooft}, where the SU($N$) gauge fields are coupled to a
{\em fundamental} Dirac fermion. Using a $1/N$ expansion, 't Hooft was
able to solve the theory in the large $N$ limit,
and its spectrum became the first to exhibit
{\em confinement}: the fermion-antifermion (mesonic) bound states
are arranged in a single linear
Regge trajectory, i.e.~the bound state masses (squared) are proportional
to an integer excitation number. The restriction to two dimensions
means that
the gauge fields (gluons) are not dynamical. Their equations of motion
show up as constraint equations on the way to deriving the Hamiltonian
of the theory. This is in stark contrast to four-dimensional
QCD, so to study the dynamics of adjoint degrees of freedom QCD$_{2A}$ was
devised. Here, the adjoint fermions form bound states analogous to
glueballs in higher dimensional pure Yang-Mills theories. Since the adjoints
are fermions, sometimes these bound states are called {\em gluinoballs}
and interpreted as closed string excitations.
Since the gluinoballs can contain an arbitrary number of (fermionic) partons,
its spectrum is much richer. Two Regge trajectories
of bound states are known to exist; from a conjectured Hagedorn transition
one expects an exponentially increasing density of states.
Furthermore, the two-index fields allow for discrete
versions of $\theta$-vacua \cite{Witten84}, i.e.~non-trivial
topological sectors. Numerical work \cite{BDK,GHK,UT1} has naturally
focused on the massive bound-state spectrum of both the massive and the
massless theory, occasionally exploring the ramifications of the massless
sector for the massive spectrum \cite{GHK} , based on the universality of
two-dimensional theories \cite{KutasovSchwimmer}.

One strand of new results stems from the finding in \cite{GHK} that
there exist multi-particle states built from the true degrees of
freedom of the theory, i.e.~the single-particle states made of strings
of adjoint fermions. The masses of these multi-particle states at a
certain harmonic resolution $K$ can often be expressed as sums of
masses at lower resolutions, such that $K_1+K_2=K$ in typical DLCQ
relations 
\beq\label{DLCQMPS}
M^2(K)= K\left[ \frac{M^2(K_1)}{K_1}+\frac{M^2(K_2)}{K_2}\right].
\eeq
However,
some such relations are only approximate, and one runs into
problems with statistics, e.g. two fermionic single-particle states
forming a fermionic multi-particle state, so one has to invoke the
massless sector for explanations. Nonetheless, the great
advantage of the DLCQ approach with its consistent IR cutoff ($1/K$
plays the role of smallest momentum), is that one generates a faithful
representation of the underlying algebraic structure, and can thus use
representation theory to classify and relate states.  This program has
recently reached near completion, when Klebanov et al.~extended
QCD$_{2A}$ by including $N_f$ fundamental Dirac fermions
\cite{Kleb21}, and considered the theory at finite $N$ \cite{Kleb22}.
The trove of results and explanations of degeneracies leaves little
doubt that the spectrum is well understood qualitatively in the
light-cone gauge $A^+=0$.  In particular, the massless theory
($m_{adj}=0$) can be ``deconstructed'' on the grounds of the Kac-Moody
algebra of its currents and a newfound $osp(1|4)$ symmetry related to
four ``supercharges'' involving all dynamic fermion fields. The
authors of \cite{Kleb21} find that degeneracies, e.g~between bosonic
and double-trace states of fermionic gluinoballs and massive fermionic
mesons, persist even when the fundamental fermions are made massive,
$y_{fund}>0$.  Therefore they reason
that the below-threshold bound states point to a attraction between
the fundamental quarks and antiquarks at short distances, while the
existence of a threshold at which a continuum starts means that the
attractive force is {\em screened} by the adjoint massless fermions.


An argument in \cite{Jacobson} corroborated in Ref.~\cite{Smilga21},
based on index theorems (``mod 2 argument'') that both massive {\em
  and} massless $QCD_{2A}$ are strongly confining\footnote{States are
  color singlets {\em and} Wilson loop has area law.} theories for odd
$N$ has recently been shown to apply only to the theory with an {\em
  additiona}l four-fermion term $[\Tr(\bar \Psi\Psi)]^2$.  In fact,
the augmented model is no longer UV finite but rather has logarithmic
flow in the UV \cite{KlebPrivate}.  The current consensus seems to be
that massless QCD$_{2A}$ proper (without the additional term) is
screening, cf.~Ref.~\cite{Seifnashri}, where it is argued that the
existence of newly found super-selection sectors of the massless
theory is evidence for its {\em screening} behavior.
Also \cite{Anton,Kleb22} show that the spectrum of QCD$_{2A}$ is
largely insensitive to $N$ (which includes $N=3$), suggesting a
screening massless theory at odd $N$.


Another issue is the appearance of vacua. For instance, in \cite{Gomis} the flow of UV operators 
is mapped for generic two-dimensional field theories, and QCD$_{2A}$ in particular. It is argued that
a subset of UV operators ends up as vacua in the extreme infrared where the theory is describable 
as a topological field theory, and thus would not show
up as massive excitations in a Hamiltonian approach. 
Due to the (naive) trivial vacuum in light-cone quantization,
one has a hard time dealing with multiple vacua in this approach. 
Most often DLCQ-based work \cite{BDK,GHK,UT1,Kleb21,Kleb22} assumes light-cone gauge to be
principally sound even in the continuum limit. 
One might be worried that gauge or fermionic zero-modes, absent 
in the discretized theory with anti-periodic boundary conditions,
will change the spectrum in the continuum limit. In fact, the typical DLCQ
(free) multi-particle spectrum, Eq.~(\ref{DLCQMPS}),
described above suggests the existence of a
discrete vacuum angle analogous to a $\theta$-vacuum. This was worked out
by the authors of Ref.~\cite{CartorRobPin97} for SU(2).
Their set of vacuum states\footnote{While SU(2) is not generic 
  in some respects \cite{Jacobson}, a set with similar properties
  has to exist for $N>2$.}
would imply copies of the masses of
lower-resolution versions of the theory consistent with Eq.~(\ref{DLCQMPS}).
If this interpretation is correct, then the spectrum
of QCD$_{2A}$ should look much cleaner in the continuum limit, where
artifacts of regularization are absent.
Additionally, one can ask how to get rid of the ``threshold
bound states'', if they are not genuine content of the
theory. As shown in \cite{Kleb21} from representation theory, these
single-trace states are degenerate with multi-trace states of the
adjoint theory (gluinoballs),
and with multi-string states if fundamental fermions are
added and meson (open string) states appear. As mentioned, bosonization
projects out only the {\em exact} multi-particle states.
It is unclear what the role of the {\em approximate} multi-particle states
is. It could be that any discrete approach
fails: these states have to be present at any finite representation
of the underlying Kac-Moody algebra, but not necessarily in the continuum limit.
The crux is that precisely the regularization makes it possible to have a
faithful representation of the algebra in the first place; the continuum
limit is the infinite parton limit, regardless of whether fermions or currents
are used. 

In light of these results, we hope that the present
work via a different, basis function approach illuminates the theory
in a complementary way. Instead of using the underlying algebraic structure,
we tackle the theory via the constraints of the integral equation inherent
in its Hamiltonian.
We use the {\em asymptotic} spectrum of the theory \cite{UT4}
to evaluate the spectrum of the {\em full} theory.  This allows
us to understand the
masses of the low-lying spectrum, and the emergence of the
multi-particle states in the full theory as described by the
asymptotic degrees of freedom. Our work shares the
shortcomings of a fermionic theory with its predecessors, namely the
appearance of exact multi-particle states. On the other hand,
we will establish that discretization {\em per se}
is no fundamental flaw. While quantitatively we do not completely agree
with previous results, we see no evidence that the continuum limit would
qualitatively change conclusions about the spectrum. For instance, one
might be worried that today's feasible resolutions of DLCQ approaches, while
substantially better than decades ago, are still too modest to tame the
divergencies of the theory enough to render results accurate. 

The paper is organized as follows. In Section 2 we describe the theory
in standard Lagrangian form and highlight the challenges in solving
the associated eigenvalue problem.  In Sec.~3 we take a look at the
asymptotic theory and its relation to the method of exhaustively-symmetrized
light-cone quantization (eLCQ) \cite{UT4}. This leads in Sec. 4 to the
construction of a complete set of (asymptotic) eigenfunctions, which
agree well with known solutions of the theory.  These basis functions
are then used in Sec.~5 to solve the {\em full} theory. To do so, we
have to generalize the procedure to integrate over the Coulomb
divergence to be applicable to adjoints. The
evaluation of the Hamiltonian matrix elements is quite involved
in the basis at hand. Nonetheless, it can be done
completely analytically, so that the result for a fixed parton number
is free of all
singularities and approximations --- an important feature of the approach.
We therefore describe the
calculation in some detail, but do so in the appendix, as the main
focus is on the {\em spectrum} of the theory, which we display in
Sec. 6 and then discuss in Sec.~7 before concluding. We note
preemptively, that our work is {\em complementary} to other approaches
\cite{GHK, Katz}. The hope is to get a clearer picture via a view of
the theory from a different angle.
While we cast doubt on the accuracy of other
approaches and we cannot do much better (outside of the asymptotic
regime) because of eLCQ's severe limitation on parton number, we
believe we can show that discretized or compactified approaches are {\em not}
fundamentally flawed when applied to QCD$_{2A}$.

\section{Theory Overview}
\label{StandardDescription}

\subsection{Lagrangian and Mode Expansion}

QCD$_{2A}$ is a non-abelian Yang-Mills theory in two dimensions 
coupled to fermions in the adjoint representation. It is
based on the Lagrangian
\beq\label{Lagrangian}
{\cal L}=Tr[-\frac{1}{4g^2}F_{\mu\nu}F^{\mu\nu}+
i\bar{\Psi}\gamma_{\mu}D^{\mu}\Psi],
\eeq
where $\Psi=2^{-1/4}({\psi \atop \chi})$, 
with $\psi$ and $\chi$ 
being $N\times N$ matrices. The field strength is
$F_{\mu\nu}=\partial_{\mu}A_{\nu}-\partial_{\nu}A_{\mu}+i[A_{\mu},A_{\nu}]$,
and the covariant derivative is defined as $D_{\mu}=\partial_{\mu}
+i[A_{\mu},\cdot]$.
We will use
light-cone coordinates $x^\pm=(x^0\pm x^1)/\sqrt{2}$,
where $x^+$ plays the role of a time and work in the light-cone gauge,
$A^+=0$, as is customary --- but see Sec.~\ref{SecIntro} and
Ref.~\cite{CartorRobPin97}.
The theory has been discussed in the literature for a while,
so we refer the reader to Refs.~\cite{DalleyKlebanov, BDK, GHK, Kleb21}
for further details. 

The dynamics of a system of adjoint fermions interacting via
a non-dynamical gluon field in two dimensions are described by the
light-cone momentum operator $P^+$ and the Hamiltonian operator $P^-$.
Following the canonical procedure
involving the energy-momentum tensor $\Theta^{\pm\pm}$,
we express the operators in terms of the dynamic fields, namely the
right-moving adjoint fermions $\psi_{ij}$ quantized by imposing anticommutation 
relations at equal light-cone times
\beq\label{PhiCR}
\left\{\psi_{ij}(x^{-}), \psi_{kl}(y^{-})\right\} = \frac{1}{2}\,
\delta(x^{-}-y^{-})\bigl ( \delta_{il} \delta_{jk}-
{1\over N}\delta_{ij} \delta_{kl}\bigr).
\eeq
The operators are then
\beas
P^+&=&\int dx^-\Theta^{++}=\frac{i}{2}\int dx^- \Tr\{\psi \partial_-\psi\}\\
P^-&=&\int dx^-\Theta^{+-}=-\frac{1}{2}\int dx^- \Tr\{im^2\psi
\frac{1}{\partial_-}\psi+g^2 J^+\frac{1}{\partial^2_-}J^+\},
\eeas
where we introduced the right-moving components $J^+_{ij}=\psi_{ik}\psi_{kj}$
of the SU(N) current.
One uses the usual decomposition of the fields in 
terms of fermion operators
\beq\label{PhiExpansion}
\psi_{ij}(x^-) = {1\over 2\sqrt\pi} \int_{0}^{\infty} dk^{+}
\left(b_{ij}(k^{+}){\rm e}^{-ik^{+}x^{-}} +
b_{ji}^{\dagger}(k^{+}){\rm e}^{ik^{+}x^{-}}\right ),
\eeq
with anti-commutation relations following from Eq.~(\ref{PhiCR})
\beq\label{Commy}
\{b_{ij}(k^{+}), b_{lk}^{\dagger}(p^{+})\} =
\delta(k^{+} - {p}^{+})
(\delta_{il} \delta_{jk}-\frac{1}{N}\delta_{ij} \delta_{kl})\,\,.
\eeq
The dynamics operators then read
\begin{eqnarray}\label{ModeDecomp}
P^+ &=& \int_{0}^{\infty} dk\ k\, b_{ij}^{\dagger}(k)b_{ij}(k)\ ,\\
P^{-} &=& {m^2\over 2}\, \int_{0}^{\infty}\nonumber
{dk\over k} b_{ij}^{\dagger}(k)
b_{ij}(k) +{g^2 N\over \pi} \int_{0}^{\infty} {dk\over k}\
C(k) b_{ij}^{\dagger}(k)b_{ij}(k) \\
&&+ {g^2\over 2\pi} \int_{0}^{\infty} dk_{1} dk_{2} dk_{3} dk_{4}
\biggl\{ B(k_i) \delta(k_{1} + k_{2} +k_{3} -k_{4})\nonumber \\
&&\qquad\qquad\times(b_{kj}^{\dagger}(k_{4})b_{kl}(k_{1})b_{li}(k_{2})
b_{ij}(k_{3})-
b_{kj}^{\dagger}(k_{1})b_{jl}^{\dagger}(k_{2})
b_{li}^{\dagger}(k_{3})b_{ki}(k_{4})) \nonumber\\
&&\qquad + A(k_i) \delta (k_{1}+k_{2}-k_{3}-k_{4})
b_{kj}^{\dagger}(k_{3})b_{ji}^{\dagger}(k_{4})b_{kl}(k_{1})b_{li}(k_{2})
 \nonumber\\
&&\qquad + \frac{1}{2} D(k_i) \delta (k_{1}+k_{2}-k_{3}-k_{4})
b_{ij}^{\dagger}(k_{3})b_{kl}^{\dagger}(k_{4})b_{il}(k_{1})b_{kj}(k_{2})
\biggl\}, \nonumber
\end{eqnarray}
with
\begin{eqnarray}
A(k_i)&=& {1\over (k_{4}-k_{2})^2 } -
{1\over (k_{1}+k_{2})^2}\ , \label {A} \\
B(k_i)&=& {1\over (k_{2}+k_{3})^2 } - {1\over (k_{1}+k_{2})^2 }, \nonumber\\
C(k)&=& \int_{0}^{k} dp \,\,{k\over (p-k)^2},\nonumber \\ 
D(k_i)&=& \frac{1}{(k_{1}-k_{4})^2} - \frac{1}{(k_{2}-k_{4})^2}\nonumber,
\end{eqnarray}
where the trace-splitting term $D(k_i)$ can be omitted at large $N$, and 
the parton-number violating term is proportional to $B(k_i)$. 
The structure of the Hamiltonian $P^-$ is 
\beq\label{StructureOfHamiltonian}
P^-= P^-_{m}+P^-_{ren}+ P^-_{PC,s}+P^-_{PC,ns}+P^-_{PV} + P^-_{finite\, N}.
\eeq
While the mass term $P^-_{m}$ is absent in the massless theory, the 
associated renormalization operator $P^-_{ren}$ needs to be included. 
Parton-number violating terms, $P^-_{PV}$, couple blocks of 
different parton number. Parton-number conserving interactions 
$P^-_{PC}$ are block diagonal, and may include singular($s$) or
non-singular($ns$) functions of the parton momenta.
The finite $N$ term $P^-_{finite\, N}$ is proportional to $D(k_i)$.
For details see \cite{DalleyKlebanov,BDK,UT3}.

\subsection{The Eigenvalue Problem}

To find the spectrum of the general Hamiltonian (\ref{StructureOfHamiltonian})
we solve the eigenvalue problem
\beq\label{EVP}
2\hat{P}^+\hat{P}^-|\Psi\rangle\equiv \hat{H}_{LC}|\Psi\rangle=M^2|\Psi\rangle
\eeq
where $H_{LC}$ is the so-called {\em light-cone Hamiltonian}.
This is a daunting task, since its eigenkets $|\Psi\rangle$ will be linear
combinations\footnote{Summation is over even(odd) integers and starts at 2(3) in the bosonic(fermionic) sectors of the theory.}
$
|\Psi\rangle=\sum_{r}^{\infty}|\Psi_r\rangle
$
of states of definite (fermion) parton number $r$
{\small
\beq\label{TheStates}
|\Psi_r\rangle=
\int_0^{\frac{1}{r}} dx_1
\left(
\prod^{r-1}_{j=2}
\int^{1-(r-j+1)x_1-\sum^{j-1}_{k=2}x_k}_{x_1}dx_j
\right)
\frac{\psi_r(x_1,x_2,\ldots,x_r)}{N^{r/2}}
Tr[b(-x_1)\cdots b(-x_r)]|0\rangle.
\eeq
}
In practice, one transforms the eigenvalue problem, Eq.~(\ref{EVP}),
into an integral equation for the {\em wavefunctions}
$\psi_r(x_1,x_2,\ldots,x_r)$ which distribute momentum between the partons
and the total momentum is set to unity by Lorentz invariance.
For instance, in the two-parton sector we have the 't Hooft-like equation
\beq\label{tHooftEqn}
\frac{m^2\psi_2(x)}{x(1-x)}-\frac{2g^2N}{\pi}\dashint_0^1\frac{\psi_2(y)}{(x-y)^2}dy=\frac{\tilde{m}^2\psi_2(x)}{x(1-x)}+\frac{2g^2N}{\pi}\dashint_0^1\frac{\psi_2(x)-\psi_2(y)}{(x-y)^2}dy=M^2\psi_2(x),
\eeq
where $\tilde{m}$ is the renormalized fermion mass and $\psi_2(x_1,x_2)=\psi_2(x,1-x)\equiv \psi_2(x)$ due to the
momentum constraint
\beq\label{MomentumConstraint}
P^+=\sum_{i=1}^r x_i=1,
\eeq
which defines the {\em physical Hilbert space}, i.e.~the hyperplane of the naive Hilbert space $[0,1]^r$ on which
this constraint is realized.

A good understanding of the integration domain and therefore the  
 Hilbert space will help us to crucially simplify evaluations
of matrix elements
for the numerical work in Sec.~\ref{SecNumericalSolution}.
The quantum theory consists of states with certain symmetries, such as
cyclic permutations due traced fermionic operators. Since states related by  
symmetry operations are equivalent,
only a small region of the naive Hilbert space
is necessary to completely and uniquely describe the dynamics of the
physical system. We call such a region a {\em unique Hilbert space cell} (uHS).
Without loss of generality, we single out the first $r-1$ momenta, i.e.~project 
onto the $(x_1, x_2,\ldots, x_{r-1})$ hyperplane. 
The uHS is then constructed by ordering tuples of momentum fractions $x_i$
so that $x_1$ is the smallest, $x_2$ the second smallest, etc.,
i.e.~parton momenta are
monotonically increasing as possible\footnote{Only cyclic reordering
  is allowed, so minimizing $x_1$ may preclude
  $x_2$ minimization, etc..},
which leads to $x_1\le 1/r$, and
\beq\label{UniqueHS}
{\rm uHS}\doteq\left[0,\frac{1}{r}\right]\times \prod_{j=2}^{r-1}
\left[x_1,1-(r-j+1)x_1-\sum_{k=2}^{j-1}x_k\right].
\eeq
In fact, we have already written out the resulting prescription for integral limits 
in Eq.~(\ref{TheStates}).
The upper integral limits, e.g.~in the unique Hilbert space cell volume
\beq\label{HSVolumeCorrect}
{\cal N}_r=\int_0^{1/r} dx_1\left(\prod^{r-1}_{j=2}
\int^{1-(r-j+1)x_1-\sum^{j-1}_{k=2}x_k}_{x_1}dx_j\right)=\frac{1}{r!}
\eeq
monotonically {\em decrease}. Yet, the last momentum fraction, $x_{r-1}$,
will be the largest, since all redundant cyclic permutations will be purged,
and the tuple $(0,0,\ldots, 0,1-\sum_j x_j)$ is the first to be constructed in
Fock-space generating algorithms such as DLCQ.
Note that $(1/r,1/r,\ldots, 1/r)$ is a unique point, and $x_1$ is a
special variable with this choice of ordering. Of course, the
  formalism is symmetric in all $x_i$, but it is not
  practical to formulate the approach such that the symmetry is {\em
    manifest}.

A physical way to think of the integral domain boundaries is as planes of
equal momenta, $x_i=x_j$, where particle identities,
i.e.~longitudinal momenta, get swapped. 
For instance, $x_3=x_1$ implies $x_2=1-x_1$ for three partons.
Pauli exclusion then dictates that wavefunctions be extreme or
zero at these boundaries.

\section{The Asymptotic Theory and eLCQ}
\label{SecAsymptoticTheory}

The coupling of parton sectors is due to 
the parton-number violating interaction $P^-_{PV}$, i.e.~pair
creation. This coupling suggests that we have to solve the problem on all
scales. However, things simplify in the large $M^2$ limit, since the form of the
integral equation
allows one to separate the long-range Coulomb interaction from the
rest, which apart from pair creation consists of the non-singular (regular)
parton-diagonal interactions. The main argument is that for large eigenvalues
$M^2$ one can neglect the mass term proportional to $m^2$ and the
behavior of the integral away from the pole at $x=y$. This allows us to push
the integration limits to infinity and to replace Eq.~(\ref{tHooftEqn}) with
\[
-\frac{2g^2N}{\pi}\int_{-\infty}^{\infty}\frac{\phi(y)}{(x-y)^2}dy=M^2\phi(x),
\]
which has sinusoidal eigenfunctions\footnote{In the massless sectors, regular (non-singular) interactions can lead to large 
corrections via singularities at the ``boundary''. Alas, the eigenfunctions are still sinusoidal.}. We renamed
the wavefunction to $\phi$ to make it clear that this is an approximation, and will
use $\Psi(x)$ and $\psi(x)$ for the full theory.
The Coulomb problem itself is intricate for adjoint fermions due to
the ``connectedness'' of operators under the color trace.
Fortunately, this Gordian knot is resolvable by an exponential ansatz
paired with group theory in the guise of an exhaustive
symmetrization. The approach called {\em eLCQ} has been brought forth
in \cite{UT4}, albeit in a clumsy way, so we'll review its salient
features in a streamlined fashion here.
In short, the approach is based the fact that the total momentum can be set to
unity by Lorentz invariance plus the following five defining characteristics
of QCD$_{2A}$:
\begin{enumerate}
\item In the asymptotic regime, the simplest solution of the integral equation
  based on the singular Coulomb interaction (inverse square behavior) is pure
  phase
  \[
  \int ^{\infty}_{-\infty} \frac{e^{i\pi n y}}{(x-y)^2}dy = - |n| \pi^2 e^{i\pi n x}.
  \]
  Therefore a reasonable ansatz for the generic solution is
  \beq\label{SimpleAnsatz}
  \chi_{\vec{\nu}}(x_1,\ldots, x_r) = \prod_{j=1}^{r-1}e^{i\pi n_j x_j}=e^{i\pi \sum_{j=1}^{r-1}n_j x_j}\equiv e^{i\pi \vec{\nu}\cdot\vec{x}},
  \eeq
  where the constraint momentum compels us to operate on a $(r-1)$-hyperplane
  of the naive $r$-dimensional Hilbert space of $r$ parton states, which is signified by the {\em Greek} letter.
 An eigenfunction of the Hamiltonian will then be a sum of such
  exponential terms, written for now symbolically as  
  \[
  \phi_{\vec{\nu}}(\vec{x})=\phi_{\vec{\nu}}(x_1,\ldots, x_r) = \sum_{{\cal G}\vec{\nu}} (-1)^{s(\vec{\nu})} e^{i\pi \vec{\nu}\cdot\vec{x}}=\sum_{{\cal G}\vec{\nu}} (-1)^{s(\vec{\nu})}\chi_{\vec{\nu}},
  \]
  where the sign of the term is determined via $s(\vec{\nu})$
  by group theoretical arguments, and the sum is over all permutations $g\in{\cal G}$ allowed
  by the symmetries of the theory, see Eq.~(\ref{TheGroup}), below.
\item The cyclic structure of the Hamiltonian
  $\hat{P}^-$ results in a mandatory (anti-)symme\-trization of its
  eigenfunctions under
  \[
  {\cal C}: (x_1,x_2,\ldots,x_r)\longrightarrow (x_2,x_3,\ldots,x_r,x_1).
  \]
  In particular,
  \beq\label{Cyclic}
  \phi(x_1,x_2,\ldots,x_r)=(-1)^{(r+1)}\phi(x_2,x_3,\ldots,x_r,x_1).
  \eeq
\item The reality of the fermions 
  results in a mandatory (anti-)symmetrization of the eigenfunctions under
  complex conjugation. One way of achieving this is to ``invert'' the momenta
  with
  \[
  {\cal I}: (x_1,x_2,\ldots,x_r)\longrightarrow (1-x_1,1-x_2,\ldots,1-x_r).
  \]
\item The vanishing of the massive wavefunctions if one momentum fraction is zero $\phi(0,x_2,x_3,\ldots,x_r)=0$ and the vanishing of the derivative
  $\frac{\partial \phi}{\partial x_1}(0,x_2,x_3,\ldots, x_r)=0$ of the massless
  wavefunctions
 results in a mandatory (anti-)symmetrization of the eigenfunctions under the so-called 
 {\em lower-dimensional inversion}
\beq\label{Sx}
{\cal S}: (x_1,\ldots x_r)\rightarrow (x_1,1-x_2-x_1, 1-x_3,1-x_4,\ldots,1-x_r-x_1).
\eeq
\item Finally, the symmetry of the Hamiltonian under flipping of color indices
  \[
  {\cal T}: b_{ij}\longrightarrow b_{ji}
  \]
  {\em can} be used to (anti-)symmetrize under
  \[
  {\cal T}: (x_1,x_2,\ldots,x_r)\longrightarrow (x_r,x_{r-1},\ldots,x_1).
  \]
Note that one has to distinguish the transformation of the {\em state} under ${\cal T}$
(sector identifier $T_{state}$) from the behavior of its {\em wavefunction} under momentum reversal
(symmetry quantum number $T$). For instance, 
in the literature \cite{GHK,UT1},  the lowest fermionic state with
$M^2\approx 5.7\frac{g^2N}{\pi}$ has been labeled as a ${\cal T}$ even
state. While indeed its wavefunction is constant and trivially even
under reversal of the momenta,
the state does lie in the $T_{state}=-1$ sector of the theory. 
Sectors are defined by the states', not the
wavefunctions' symmetry properties.  
\end{enumerate}
These five automorphisms of momentum space
completely determine the asymptotic eigenfunctions.
It is often useful to think of them as operating in excitation number space:
\begin{eqnarray}
  {\cal C}: (n_1,n_2,\ldots,n_{r-1})&\longrightarrow& (-1)^{n_{r-1}}(-n_{r-1},n_1-n_{r-1},\ldots,n_{r-2}-n_{r-1}),\\
  {\cal I}: (n_1,n_2,\ldots,n_{r-1})&\longrightarrow& (-n_1,-n_2,\ldots,-n_{r-1}),
  \nonumber\\
  {\cal S}: (n_1,n_2,\ldots,n_{r-1})&\longrightarrow& (n_1-n_2,-n_2,-n_3,\ldots,-n_{r-1}),\nonumber\\
  {\cal T}: (n_1,n_2,\ldots,n_{r-1})&\longrightarrow& (-1)^{n_1}(-n_1,n_{r-1}-n_1,n_{r-2}-n_1,\ldots,n_2-n_1).\nonumber
\end{eqnarray}
One constructs the eigenfunctions by acting with the operators on a
generic single-particle momentum state
\beq\label{SPstates}
|\vec{x}\rangle=|x_1,\ldots x_r\rangle=\Tr\left\{b^{\dagger}(x_1)\cdots b^{\dagger}(x_r)\right\}|0\rangle
\eeq
or equivalently on the function
\[
\chi_{\vec{\nu}}(x_1,x_2,\ldots,x_r)=
\langle x_1,x_2,\ldots, x_r|n_1,n_2,\ldots,n_{r-1}\rangle
=e^{i\pi \vec{\nu}\cdot\vec{x}},
\]
to produce all distinct operators (algebraic {\em words})
$
{\cal O}_k={\cal C}{\cal I}\cdots{\cal S}{\cal T}{\cal C}^2\cdots{\cal I}
    {\cal S},$
where $k$ is an enumerative index.
The (complete) set of such operators is called the total symmetrization group
\beq\label{TheGroup}
{\cal G}\equiv\{{\cal O}_k\}.
\eeq
We find that the order of ${\cal G}$ is $2r!$, which is twice the order of the
symmetric group $S_r$. This makes sense, since we are in essence
permuting $r$ objects (the fermion operators) with an additional
optional $Z_2$ symmetrization with respect to ${\cal T}$.
In essence,
the exchange of momentum is the same as swapping partons since they are
identified by their longitudinal momentum. Even if we are limited to
transpositions of adjacent operators, we can still cover the entire symmetric
group $S_r$, albeit with an algebraic structure that reflects this. It is
unclear whether this scheme can be generalized to higher dimensions. 

In practice, one uses a computer to construct these states. Note
that this has to be done symbolically since we need to decide which operators
are distinct, because numerical matching does not
suffice\footnote{For example, $n_1+n_2\neq n_2-n_3$, but for $\vec{n}=(1,2,-1)$ we
  have $3=3$.}. 
The direct product of inversions, ${\cal I}$, reorientations ${\cal T}$, and
cyclic permutations ${\cal C}$ forms a subgroup ${\cal B}$ of order $4r$
of the full group ${\cal G}$ of order $2r!$. This means we can organize
the ``statelets'' $|n_1,n_2\ldots, n_{r-1}\rangle$ as right cosets
of  ${\cal B}$ in ${\cal G}$ where the operators in ${\cal B}$ act
on $\frac{1}{2}(r-1)!-1$ operators collected in the {\em exhaustive set}
\[
  {\cal E}=\{{\cal S}_1, {\cal S}_2,\ldots, {\cal S}_{(r-1)!/2-1}\}
\]
plus the identity. In \cite{UT4} we tried to classify these ${\cal S}$ operators
further, but it suffices to view them as composite operators
involving at least one instance of the ``fundamental'' ${\cal S}$ operator.
In particular, ${\cal S}_1={\cal S}$. Then the ${\cal S}_i$ are algebraic words
like ${\cal SICTSC}^2$, and for our purposes we only need to know
the number of times the letter ${\cal S}$ appears in the word
(two in the example) which we refer to as
{\em S-ness}. The latter determines the sign of the statelet within the
eigenstate. Since ${\cal S}$ is a $Z_2$ operator with eigenvalues $S=\pm 1$,
  we get a minus sign only for odd $S$-ness in the $S=-1$ sectors of the
  theory.

We conclude that the vanishing of the wavefunction whenever $x_i=0$ is
rather tricky to implement.
Indeed, to achieve $\phi(x_i=0)=0$ we need to subtract a term
$e^{i\pi \vec{\nu}\cdot\vec{x}}$ that is identical to an existing term at
$x_i=0$ but different otherwise. But this new term then becomes part of
the problem: it too needs to be canceled by yet another term, and so
on until the possibilities to form distinct terms are {\em exhausted}.
On the other hand, once the eigenfunctions are constructed reflecting the
symmetries and structure of the Hamiltonian,
they will automatically respect Pauli exclusion
(vanish for $x_i=x_f$ where necessary) and be (anti-)periodic at the
boundaries of the unique Hilbert space.

We now have a blueprint for the states of the theory. In excitation
space we may write it as ($|r\rangle=|n_1,n_2,\ldots, n_{r-1}\rangle$)
\bea
|\phi_{\vec{\nu}}\rangle &=& |\vec{\nu}\rangle = {\cal G}|r\rangle=|r\rangle + {\cal C}|r\rangle + {\cal C}^2|r\rangle +\ldots\nonumber\\
&=&|n_1,n_2,\ldots, n_{r-1}\rangle+(-1)^{n_{r-1}}|-n_{r-1},n_1-n_{r-1},\ldots, n_{r-2}-n_{r-1}\rangle +\ldots.\nonumber
\eea
The next step is to fill in integers for the excitation numbers $n_i$
and check which combinations lead to viable states\footnote{The $n_i$
  can be even, odd, positive, negative
  or zero, and may lead to vanishing or redundant wavefunctions
  \cite{UT3, UT4}. 
  }.
The task to generate these {\em bona fide} states is best left to a
computer. The results are displayed in Table \ref{StateTable}.

\begin{table}
  \begin{footnotesize}
    \centerline{
\begin{tabular}{|c|lc|l|l|}\hline      
  \rule[-3mm]{0mm}{8mm}
  $r$& T I S &Sector$^{\rm mass}_{T_{state}}$ & Excitation numbers of lowest states & Masses ($g^2N\pi$)\\\hline
2 & $-+$ & $|o\rangle^0_+$ & $(1),(3),(5),(7)$ & $2,6,10,14$\\
  & $--$ & $|e\rangle^{\mu}_+$ & $(2),(4),(6),(8)$ & $4,8,12,16$\\
\hline
3 & $--$  & $|ee\rangle^{0}_+$ & $(4,2),(6,2),(8,2),(8,4)$ & $8,12,16,16$\\
 & $++$  & $|ee\rangle^{0}_-$ & $(0,0),(2,2),(4,2),(4,0),(6,2)$ & $0,4,8,8,12$\\
  & $-+$  & $|ee\rangle^{\mu}_+$ & $(6,2),(8,2),(10,4),(10,2)$ & $12,16,20,20$\\
  & $+-$  & $|ee\rangle^{\mu}_-$ & $(2,0),(4,0),(6,2),(6,0)$ & $4,8,12,12$\\
\hline
4 & $+--$  & $|oeo\rangle^{0}_+$ & $(3,2,1),(5,4,3),(5,6,3),(7,6,3)$ & $6,10,12,14$\\
  & $-++$  & $|oeo\rangle^{0}_-$ & $(1,2,1),(3,2,1), (3,4,3),(3,6,3)$ & $4,6,8,10$\\
  & $+-+$  & $|eee\rangle^{\mu}_+$ & $(6,6,4),(8,8,6),(8,10,6), (10,10,6), (10,10,8)$ & $12,16,20,20,20$\\
  & $-+-$  & $|eee\rangle^{\mu}_-$ & $(4,6,4),(6,6,4), (6,8,6), (6,10,6)$ & $12,12,16,16$\\
\hline
5 & $+++$  & $|eeee\rangle^{0}_+$ & $(0,0,0,0),(2,2,2,2),(2,4,4,4),(4,4,4,2),(4,4,4,4)$ & $0,4,8,8,8$\\
  & $---$  & $|eeee\rangle^{0}_-$ & $(4,4,4,2),(4,6,6,6),(4,6,4,2),(6,6,4,2),(4,8,8,8)$ & $8,12,12,12,16$\\
  & $++-$  & $|eeee\rangle^{\mu}_+$ & $(4,6,6,4),(6,8,8,6),(6,10,10,6),(8,10,10,6)$ & $12,16,20,20$\\
  & $--+$  & $|eeee\rangle^{\mu}_-$ & $(8,10,10,6),(8,12,10,6)$,(8,14,12,8) & $20,24,28$\\
\hline
6 & $-++$  & $|oeoeo\rangle^{0}_+$ & $(1,2,3,2,1),(1,2,3,4,3),(5,4,3,2,1),(3,4,5,4,3)$ & $6,8,10,10$\\
& $+--$  & $|oeoeo\rangle^{0}_-$ & $(1,2,3,4,3),(5,4,3,2,1),(3,6,5,4,3),(5,6,5,4,3)$
& $8,10,12^{(2)}$\\
& $--+$  & $|eeeee\rangle^{\mu}_+$ & $(6,10,12,10,6),(6,10,12,12,8),(8,12,14,12,8),$
& $24^{(2)},28^{(2)},32,36$\\
&&&$(8,14,14,12,8),(8,14,16,14,8),(8,14,18,14,8)$ & \\
& $++-$  & $|eeeee\rangle^{\mu}_-$ & $(8,12,12,10,6),(10,12,12,10,6),(8,12,14,14,10),$
& $24^{(2)},28^{(3)}$\\
&&&$(8,14,14,12,8),(12,14,14,12,8)$ & \\
\hline
7 & $---$  & $|eeeeee\rangle^{0}_+$ & $(2,4,4,4,4,4),(4,6,6,6,4,2),
(4,6,8,6,4,2),$
& $8,12,16,16$\\
& & & $(4,6,8,8,8,4)$&\\
 & $+++$  & $|eeeeee\rangle^{0}_-$ & (0,0,0,0,0,0),(2,2,2,2,2,2),(2,4,4,4,4,2),
(2,4,4,4,4,4)
%
 & $0,4,8,8$ \\
  & $-+-$  & $|eeeeee\rangle^{\mu}_+$ & $(8,14,16,16,14,10),
(8,14,18,18,16,10),$ & $32,36^{(3)}$\\
& & & $(8,14,18,18,16,12),(10,16,18,18,16,12)$ &\\
  & $+-+$  & $|eeeeee\rangle^{\mu}_-$ & $(6,10,12,12,10,6),
    (8,14,16,16,14,8),
 $ & $24,32,32,36$\\
& & & $(8,14,16,16,14,10),(8,14,18,18,14,8)$&\\
\hline
8 & $+--$  & $|oeoeoeo\rangle^{0}_+$ & $(3,4,5,4,3,2,1),(5,4,5,4,3,2,1),
(3,6,5,4,3,2,1)$ & $10,12,12$\\
& $-++$  & $|oeoeoeo\rangle^{0}_-$ & $(1,2,3,4,3,2,1),(3,4,5,4,3,2,1),
(3,4,5,6,5,4,3)$ & $8,10,12$\\
& $+-+$  & $|eeeeeee\rangle^{\mu}_+$ & $(10,16,20,20,18,14,8),
(10,18,20,20,18,14,8)$ & $40,40$\\
& $-+-$  & $|eeeeeee\rangle^{\mu}_-$ & $(8,14,18,20,18,14,8),
(10,16,20,20,18,14,8)$ & $40,40$\\
\hline
9 & $+++$  & $|eeeeeeee\rangle^{0}_+$ & $(0)^8,(2)^8,(4,4,4,4,4,4,4,4),$& $0,4,8,8,8$ \\
&&&$(4,4,4,4,4,4,4,2),(2,4,4,4,4,4,4,2)$ & \\
& $---$  & $|eeeeeeee\rangle^{0}_-$ & $(4,4,4,4,4,4,4,2)$ & $8$\\
& $++-$  & $|eeeeeeee\rangle^{\mu}_+$ & $(8,14,18,20,18,14,8)$ & $50$\\
& $--+$  & $|eeeeeeee\rangle^{\mu}_-$ & $(12,18,20,22,20,18,10)$ & $44$\\
\hline
\end{tabular}}
\end{footnotesize}
  \caption{    
  The lowest states in the lowest asymptotic parton sectors
  including their quantum numbers $TIS$.
  The sectors are labeled with a subscript
  indicating behavior under ${\cal T}$ ($_{\pm}$), i.e.~$T_{state}=\pm 1$,
  and superscripts signifying massless ($^0$) and massive fermions ($^{\mu}$).
  \label{StateTable}
  }
\end{table}


\subsection{Grand Bulk Equivalence and Orthonormality}
\label{SubSecGBE}

In light of the symmetries the structure of the Hilbert space becomes
clear.  Namely, in a unique Hilbert space cell one Fock state
$\Tr\{b(-x_1)\cdots b(-x_r)\}|0\rangle$ with a specific tuple of
momentum fractions represents $2r!$ such states 
located in
$r!$ different cells\footnote{There are $r!$ cells since
  the action of ${\cal T}$ yields a state of different orientation in
  the same cell.}  of the physical Hilbert space, which is thus
tessellated.
As a consequence eigenfunctions that are orthogonal on a unique Hilbert
space cell will be orthogonal everywhere.
The $r!$ unique Hilbert space cells
of the 
physical hyperplane are connected by the four 
automorphisms ${\cal C}$, ${\cal T}$, ${\cal I}$, and ${\cal S}$ that
generate the group
${\cal G}\equiv\langle{\cal C},{\cal T},{\cal I},{\cal S}\rangle$.

Now, the symmetrization of the wavefunctions under
$\cal G$ is equivalent to putting a representative of each cell into the
linear combination in one specific cell.
Mathematically, it is the simple fact that
$({g}\nu)\cdot\vec{x}=\nu\cdot({g}\vec{x})$ up to a sign for ${g}\in{\cal G}$.
  For instance, at $r=4$ we have with ${g}={\cal C}$
  \[
  (-n_3,n_1-n_3,n_2-n_3)\cdot(x_1,x_2,x_3)=(n_1,n_2,n_3)\cdot(x_2,x_3,x_4),
  \]
with $x_4=1-x_1-x_2-x_3$ on the physical hyperplane.
In other words, a term $e^{i\pi \vec{\nu}\cdot({g}\vec{x})}$ in the
Hilbert cell ${g}|\vec{x}\rangle$ is substituting for 
the term $e^{i\pi ({g}\vec{\nu})\cdot\vec{x}}$ in the original cell.
Note that the former has the original excitation numbers. 
Therefore, integrating a $2r!$ term eigenfunction over one unique
cell is equivalent to integrating a one-term function over the entire
physical Hilbert space (pHS) projected onto the $(x_1,\ldots,x_{r-1})$ hyperplane\footnote{As implausible as it
  sounds! For example, one might be worried that there is {\em one}
  oscillating function everywhere with a given set of wavenumbers, so
  that the choice of $\vec{\nu}$ matters. However, a set of excitation
  numbers is interpreted as a different set in a different
  cell.}. We might call this {\em grand bulk equivalence} (GBE), since
it holds anywhere in the bulk of the Hilbert space.
In lieu of a proof, consider that the integral over the
second term (first cyclic permutation, $g={\cal C}$)
of the three-parton asymptotic
wavefunction is shifted by a change of variables $x_1'=x_2, x'_2=1-x_1-x_2$
with Jacobian $\frac{\partial(x_1,x_2)}{\partial(x_1',x_2')}=1$,
while its integral domain $uHS_2={\cal C}(uHS_1)$ is the cyclically permuted
version\footnote{Its boundaries are $x'_2=0, x'_2=x'_1,
  x_2'=\frac{1}{2}(1-x'_1)$.} of the original. To wit
\bea\label{GBEtrafo}
\int_{uHS_1}\!\!\!\! d^2\vec{x}\,\, e^{i\pi((n_1-n_2)x_1-n_1x_2)}
&=&\int_0^{1/3}\!\!\!\! dx_1\int_{x_1}^{1-2x_1} \!\!\!\!dx_2\,\, e^{i\pi((n_1-n_2)x_1-n_1x_2)}\\
&=&\int_0^{1/3}\!\!\!\! dx'_2\int_{x'_2}^{1-2x'_2} \!\!\!\!dx'_1\,\, e^{i\pi(n_1x'_1-n_2x'_2)}
=\int_{uHS_2}\!\!\!\! d^2\vec{x}\,\, e^{i\pi(n_1 x_1+n_2 x_2)},\nonumber
\eea
where we omitted the primes of the integration variables in the last step.
Similar transformations for the other $g\in{\cal G}$ lead to a full coverage
of the semi-naive Hilbert space $snHS=\cup_{j=1}^{r!}uHS_j$.
There are two
caveats.  Firstly, this only works if we {\em integrate}; the
wavefunctions themselves are, of course, not the same. Secondly, the
$2r!$ terms of an eigenfunction enter with different signs owing to
the quantum numbers of the symmetry sector. Since we are mostly
interested in evaluating scalar products between two wavefunctions of
the same sector, this is of no consequence as we are squaring the sign.

If exploiting GBE seems too good to be
true, note firstly that it is simply a consequence of the structure of the Hamiltonian. For instance, the much simpler 't Hooft Hamiltonian leads to fewer
constraints, hence larger (and fewer) Hilbert space cells, and therefore to the
same result: integration of a single sinusoidal function over the entire
semi-naive
Hilbert space suffices. Secondly, GBE is of little practical use unless the
function integrand is invariant under ${\cal G}$. Unfortunately, this
is not the case for the Hamiltonian matrix elements as they depend on specific
momenta on the physical hyperplane of the naive Hilbert space. If nothing else,
GBE allows for an effortless proof of the orthonormality of the
asymptotic eigenfunctions
\[
\langle\vec{\mu}|{\vec{\nu}}\rangle=
\int_{uHS}\!\!\!\!\!\!
d^{r-1}\vec{x}\,\,\phi^*_{\vec{\mu}}(\vec{x})\phi_{\vec{\nu}}(\vec{x})
=\!\!
\int_{pHS}\!\!\!\!d^{r-1}\vec{x}\,\,\chi^*_{\vec{\mu}}(\vec{x})\chi_{\vec{\nu}}(\vec{x})
=\prod_{j=1}^{r-1}\int^1_0 \!\! d{x_j}\,\, e^{-i\pi\vec{\mu}\cdot \vec{x}}e^{i\pi\vec{\nu}\cdot \vec{x}}=\delta_{\vec{\mu}\vec{\nu}}.
\]
Note that $\langle\vec{\mu}|{\vec{\nu}}\rangle=0$ only implies independent
states when $|{\vec{\nu}}\rangle$ and $|{\vec{\mu}}\rangle$ are in different
equivalence classes defined by $|{\vec{\nu}}\rangle\approx|{\vec{\mu}}\rangle$
if $|{\vec{\nu}}\rangle=g|{\vec{\mu}}\rangle$ for at least one $g\in{\cal G}$.
The Hilbert space is tessellated by $r!$ unique cells, while there are
$2r!$ exponential terms or $r!$ {\em statelets} representing
trigonometric functions. The normalization factor is
\beq\label{Normalization}
{\cal N}=\sqrt{\frac{r!{\cal N}_0}{\mbox{\#statelets}}},
\eeq
where for almost all eigenfunctions 
${\cal N}_{0}=2$ save for the constant $\vec{\nu}=\vec{0}$
functions (appearing in the odd parton $T_{state}=(-1)^{(r-1)/2}$ sectors)
we have ${\cal N}_{0}=1$.
The number of statelets varies due to ``accidental''
symmetries of excitation number tuplets,
see Eq.~(\ref{FermionicGroundState}) and similar.

\section{Constructing an Asymptotic Eigenfunction Basis}
\label{SecAsymptoticBasis}

In the previous section
we presented the {\em principles} of finding a set
of harmonic eigensolutions for asymptotic QCD$_{2A}$ and
listed the resulting {\em bona fide} states separately for the lowest parton sectors in
all viable symmetry sectors in Table \ref{StateTable}. 
We will now see that the approach can be applied to construct the {\em complete}
(asymptotic) spectrum of $QCD_{2A}$. The hope is that it can be generalized
to tackle other theories, too.

\subsection{Generic Algorithm based on Ground States}
\label{GroundStates}

Using Table \ref{StateTable}, it is not
hard to come up with general, all-parton-sector expressions
for the ``ground states'' of all symmetry and parton sectors, including
their masses and symmetry factors.
As can be gleaned from Table \ref{StateTable}, the expressions will differ
substantially for even and odd parton number as well as even and odd
excitation numbers.
We therefore need to consider eight different cases (fermionic
vs.~bosonic, massless vs. massive, and $T=\pm$). Masses $\bar{M}^2$
are in units $gN^2\pi$ (as in Table \ref{StateTable}); for instance
$\bar{M}^2=4\rightarrow M^2\approx 39.48\frac{g^2N}{\pi}$ in the usual
units.
\begin{enumerate}
\item {\bf The fermionic massless $T=+1$ sector}
  is the easiest to figure out.
  These are $T_{state}=(-1)^{(r-1)/2}$ {states}\footnote{We use $T_{state}$
    for the behavior of the {\em states} (wavefunction {\em and}
    trace of operators); $(\pm)$ indices signify $T$ not $T_{state}$ sectors.
  } and all symmetry quantum numbers are positive since the lowest state
  has $r!$ identical statelets which all have to enter with the same sign lest
  the state vanishes identically. It is easy to read off the lowest five states
  \bea\label{FermionicGroundState}
  |1\rangle^{0}_{f(+)}=|\bar{M}^2=0; 0^{r-1}\rangle\frac{1}{\sqrt{2r!}},&&\!\!\!\!\!\!
  |2\rangle^{0}_{f(+)}=|\bar{M}^2=4; 2^{r-1}\rangle\frac{1}{\sqrt{2(r-2)!}},\nonumber\\
  |3\rangle^{0}_{f(+)}=|\bar{M}^2=8; 4^{r-1}\rangle\frac{1}{\sqrt{2(r-2)!}},\nonumber&&\!\!\!\!\!\!
  |4\rangle^{0}_{f(+)}=|\bar{M}^2=8; 4^{r-2},2\rangle\frac{1}{\sqrt{2(r-3)!}},\nonumber\\
  |5\rangle^{0}_{f(+)}=|\bar{M}^2=8; 2,4^{r-3},2\rangle\frac{1}{\sqrt{8(r-4)!}},&&
  \eea
  where we displayed the {\em symmetry factor} of the states under the
  square root. For instance, the ground state has $2r!$ identical statelets
  in which all $r-1$ excitation numbers are zero $|0,0,\ldots, 0\rangle$.
  Note that the masses do not depend on the parton number $r$, and thus
  we expect high-parton-number sectors to contribute significantly if
  parton number violation is allowed.
\item {\bf The massless fermionic $T=-1$ sector} has $T_{state}=(-1)^{(r+1)/2}$
  states which
  have asymmetric excitation number tuples.
  The symmetry quantum numbers are opposite of the previous sector, so
  $TIS=(---)$
  \[
  |1\rangle^{0}_{f(-)}=|\bar{M}^2=8; 4^{(r-2)},2\rangle
  \frac{1}{\sqrt{2(r-3)!}}. 
  \]
 \item {\bf The massive fermionic $T=+1$ sector} with $T_{state}=(-1)^{(r-1)/2}$
  states can be symmetric in excitation numbers (the ground state is!).
  The symmetry quantum numbers are $TIS=(+,(-)^{\frac{r-1}{2}},(-)^{\frac{r+1}{2}})$
  and the lowest state is
  \[
  |1\rangle^{\mu}_{f(+)}=|\bar{M}^2=2\sum^r_{even\, j}r-j; r-1,2(r-2),3(r-3),
  \ldots, 2(r-2),r-1\rangle/\sqrt{2}. 
  \]
\item {\bf The massive fermionic $T=-1$ sector} with $T_{state}=(-1)^{(r+1)/2}$
  states. There is no symmetry here; all symmetry factors are one,
  $TIS=(-,(-)^{\frac{r+1}{2}},(-)^{\frac{r-1}{2}})$
  \[
  |1\rangle^{\mu}_{f(-)}=|\bar{M}^2=6r-10; r+3,2r,2(r+1),2(r+2),\ldots 2(r+2),
  2(r+1),2r, r+1\rangle. 
  \]
  Mass squared is $\bar{M}^2=12$ for $r=3$ and $\bar{M}^2=6r-10$ otherwise,
  so growing linearly with parton number $r$.
\item {\bf The massless bosonic $T=+1$ sector} with $T_{state}=(-1)^{r/2}$
  states are
  \[
  |1\rangle^{0}_{b(+)}=|\bar{M}^2=r+2; 3,4,5,\ldots \frac{r}{2}-1,
  \frac{r}{2}, \frac{r}{2}-1,\ldots, 2,1\rangle/\sqrt{(r/2+1)!}.
  \]
  The mass grows linearly with parton number.
 \item {\bf The massless bosonic $T=-1$ sector} with  $T_{state}=(-1)^{r/2-1}$,
  states having symmetric excitation number tuples
  \[
  |1\rangle^{0}_{b(-)}=|\bar{M}^2=\bar{r}; 1,2,3,\ldots \frac{r}{2}-1,
  \frac{r}{2}, \frac{r}{2}-1,\ldots, 2,1\rangle
  \frac{1}{\sqrt{2\prod^{r/2}_{j=2}j^2}}, 
  \]
  with $\bar{r}=r$ for $r>2$ and one for $r=2$.
  Note that mass is proportional to $r$, so high parton-number states
  should {\em not} contribute much to the lowest states of the full theory.
  Note also that due to the large symmetry factor there are only
  $2r!/2\prod^{r/2}_{j=2}j^2=6,20,70,\ldots$ independent statelets
  for $r=4,6,8,\ldots$. This is an example
  for an ``accidental'' symmetry we alluded to in \cite{UT4}. 
  
\item {\bf The massive bosonic $T=+1$ sector} with $T_{state}=(-1)^{r/2}$ states
  with no symmetries (factors all one). The lowest state is
  \[
  |1\rangle^{\mu}_{b(+)}=|\bar{M}^2=r\left(\frac{r}{2}+1\right); r+2,2r, 3r-2,\ldots,R,R,\ldots, 3(r-2),2(r-1),r\rangle, 
  \]
  with $R:=
  \frac{r}{2}\left(\frac{r}{2}+1\right)$.
\item {\bf The massive bosonic $T=-1$ sector} with $T_{state}=(-1)^{r/2-1}$
  states. A subset of them
  was constructed in \cite{Kutasov94}.
 A symmetric representative
 of the ground state is
 {\small \[
  |1\rangle^{\mu}_{b(-)}=|\bar{M}^2=2\sum^r_{even\, j}j; r,2(r-1),3(r-2),4(r-3),\ldots,R,
  \ldots,3(r-2),2(r-1),r\rangle/\sqrt{2}. 
  \]}
  Note that the mass increases with the parton number, and that a simpler
  representative of the state is
  \[
  |1'\rangle^{\mu}_{b(-)}=|\bar{M}^2=2\sum^r_{even\, j}j; r,0,r-2,0,r-4,0, \ldots,0,2
  \rangle/\sqrt{2}. 
  \]
  Since every other excitation number is zero in this statelet, it looks like
  the state can be characterized by $r/2$ excitation numbers,
  see Sec.~\ref{KnownComparison}.
  The symmetry factor suggests that this is not
  the most general state of the sector.
\end{enumerate}

Now that the ground states are known, we can construct the rest of the spectrum
by acting with the ``ladder operators''
\[
  \hat{\cal L}_i: |n_1,n_2,\ldots,n_i, \ldots ,n_{r-1}\rangle\longrightarrow
  |n_1,n_2,\ldots,n_i+2, \ldots,n_{r-1}\rangle,
\]
This will take us from the ``highest weight''
statelet of a bona fide state to {\em some} statelet of a different
bona fide state --- unless the latter does not exist in the symmetry
sector considered, in which case we act with ${\cal L}_{i+1}$, until a
desired number of bona fide states is produced. 


The astonishing fact that for every parton number only four of
eight TIS symmetry sectors give rise to states regardless of
whether only even or even and odd excitation numbers are allowed is --- of course ---
due to group theory.

\subsection{Extracting the Physics}
\label{PhysicsExtraction}

These combinatorics exercises entail some physics. We saw that
massless states likely have significant contributions from all higher
parton sectors. Additionally, a look at the $I$ quantum number reveals that
in the massless sectors, the $r$ and $r+2$ sectors sport opposite
trigonometric functions (sines vs.~cosines), whereas in
the massive sectors, the trigonometric functions
are the same. This holds in both the bosonic and fermionic sectors and
will dramatically change the importance of the parton-number violating
interaction in the massless versus massive sectors, see Sec.~\ref{SecNumericalSolution}.

Note that the wavefunctions 
are necessarily
even(odd) under cyclic rotations of their momenta in the odd(even)
parton sectors.  The question is then whether they are even or odd
under momentum order reversal\footnote{Actually, the question is more
  complicated for $r>3$ since there are $\frac{1}{2}(r-1)!-1$
  additional symmetries associated with the
  lower-dimensional inversions ${\cal S}_i$.}. For states with three fermionic
operators, asymmetric states \beq\label{Tpartners} {\cal T}|a\rangle
=|b\rangle; \quad |a\rangle \neq|b\rangle \eeq are combined with(out)
a relative sign to form ${\cal T}$-even(odd) states
$\sqrt{2}|T\pm\rangle=|a\rangle \pm|b\rangle$. Therefore we need a
${\cal T}$-even wavefunction to consistently distribute the momenta. The
scheme reverses at $r+2$, where the additional operators yield
an extra sign when putting flipped indices in order. Therefore the five-fermion
states in the same $T_{state}$ sector have asymmetric states under
momentum reversal, and hence a relative sign in
Eq.~(\ref{Tpartners}). Hence, a $\cal T$-even wavefunction is necessary.
Only the symmetric wavefunctions produce massless states,
so surprisingly a generic all-sector wavefunction of the full theory will
have opposite $\cal T$-symmetry of the wavefunctions in adjacent parton sectors.
So it will, for instance, comprise a massless asymptotic state in the $4s-1$
parton sectors, but asymptotic states with rather high mass ($\bar{M}^2=8$)
in the
$4s+1$ sectors ($s=1,2,3,\ldots$). This leads to a large mass
gap between the lowest states in a sector, explaining the purity in parton
number \cite{DalleyKlebanov,BDK} and the fact that almost any brutalization
of the theory leads to the correct mass of the ground state(s). 


In fact, much of the insight into the lowest states could have been gleaned
from earlier work, e.g.~\cite{BDK,GHK,UT1}, and becomes almost trivial
in the eLCQ approach.
For instance, the fact that the
lowest fermionic and bosonic states are isolated in mass and pure in
parton number is a result of the asymptotic spectrum
having massless states only in the fermionic sectors.
In particular,
the lowest state of the full theory is a three-parton fermion and not a
two-parton boson since the lowest bosonic
state has a mass
  (squared) of about $\bar{M}^2=1$ (${M}^2\approx 10\frac{g^2N}{\pi}$) whereas the
lowest asymptotic fermionic state
in the adjoint theory 
is massless, and acquires a mass (squared) of
about $r(r-1)\frac{g^2N}{\pi}\approx 5.7\frac{g^2N}{\pi}$ via the non-singular
parton-number preserving interaction, see Appx.~\ref{AppxRegular}.

In eLCQ it is easy to see that the lowest states are isolated, since the
higher-parton states start at quite high mass, see Fig.~\ref{eLCQvseps}. In particular,
the lowest four- and five-parton states are $M^2\approx 36\frac{g^2N}{\pi}$
and $M^2\approx 55\frac{g^2N}{\pi}$, so about three- and
ten times the ground state mass. Also the surprising find \cite{DalleyKlebanov}
that the lowest $T_{state}+$ state is a very pure five-parton state is clear in
eLCQ: the lowest three-parton asymptotic state is at
$M^2\approx 50\frac{g^2N}{\pi}$,
since\footnote{This remains true even if non-singular terms are added,
because the expectation value of the non-singular term is zero for
three-parton states in that sector.}
the lowest available excitation numbers in that
sector ($TI=--$) are $(4,2)$, implying a large mass.
But that's where it stops: the lowest six- and
  seven-parton states have a similar mass as their lower-parton
  counterparts, and therefore none of the higher states are pure in
  parton number.
 Incidentally, in Ref.~\cite{BDK} it was speculated that this isolation
  and purity of the lowest states could be related to the surprising success
  of the valence quark model in full four-dimensional QCD.
  We can also answer the question as to why
  the lowest mass grows linearly with the number of flavors \cite{Adi2000}.
  The terms in the Hamiltonian, Eq.~(26) of \cite{UT1}, are at most
    linear in $N_f$, and thus in the absence of eigenvalue repulsion due to
    isolation, the trajectory $M^2_{low}(N_f)$ is necessary linear, see Fig.~5(b)
    of \cite{UT1}.

\subsection{Comparison to Known Solutions}
\label{KnownComparison}

As a cross-check we compare the eLCQ eigensolutions with the output of 
other approaches. Two-, four-, and six-parton wavefunctions
for one of the two bosonic
${\cal T}$ sectors were presented in Ref.~\cite{Kutasov94},
Eqns.~(4.12), (4.13) and (4.15).
The solutions in \cite{Kutasov94} are orthogonal and vanish at
$x_i=0$, so they should be a subset of the {\em massive} solutions
presented in the present note.  Indeed, the lowest states $\phi^{[7]}_4(4,2)$ and $\phi^{[7]}_6(6,4,2)$
are identical with the eLCQ solutions $\varphi_{4-}(4,0,2)$ and
$\varphi_{6+}(6,0,4,0,2)$ including the masses\footnote{See \cite{UT4}
  where the massive(massless) wavefunctions
  are labeled with $\varphi$($\phi$).}. At $r=4$ the equivalency is
$\phi_4^{[7]}(n_1,n_2)=\varphi_{4-}^{eLCQ}(n_1,0,n_2)$. However, we find that
some eLCQ states (which are numerically virtually identical with 
DLCQ results \cite{UT4}) are not reproduced. This means that the 
set of states described by \cite{Kutasov94}, Eq.~(4.13), is not complete.

In the six parton sector the situation is more complicated. While some
states of \cite{Kutasov94}, Eq.~(4.15), coincide with eLCQ states, others do
not match. Since the former {\em are} orthogonal and
vanish at the boundary, they must violate some
``internal'' boundary condition, i.e.~a zero or extremal wavefunction
on hyperplanes where momentum fractions match, $x_i=x_j$.
Crossing such hyperplanes one enters a different, redundant part of the
Hilbert space.

Needless to say, the eLCQ eigenfunctions pass a numerical orthonormality check.
Note that the first four $r=4$ eigenvalues are
pairwise degenerate, and yet their wavefunctions are orthogonal. Hence, it looks like we have all
relevant symmetries taken care of.


\section{Solving the theory with a basis-function approach}
\label{SecNumericalSolution}


Now that we have a basis of asymptotic eigenstates $\{|\phi_{\vec{\mu}}\rangle\}$,
we use it to solve the {\em full} theory.
We expand the true eigenstates as linear
combinations of asymptotic eigenstates by
diagonalizing the Hamiltonian in this asymptotic basis, i.e.~by solving the
eigenvalue problem ($\hat{H}:=2\hat{P}^+\hat{P}^-$), cf.~Eq.~(\ref{EVP})
\beq\label{ABFEVP}
\langle \phi_{\vec{\mu}}|\hat{H}_{full}|\phi_{\vec{\nu}}\rangle
=M^2\langle \phi_{\vec{\mu}}|\phi_{\vec{\nu}}\rangle.
\eeq
This is a finite matrix equation when the number of basis states is cut off at
$N_{\phi}<\infty$. Convergence is typically exponential in the number of
basis states used. Recall that the parton sectors are coupled by the
pair-production interaction, so it will be convenient to limit the number
of states separately in each parton sector so that $\sum_r N_{\phi,r}=N_{\phi}$.

To compute the matrix elements in the asymptotic basis, we need the Hamiltonian
in a basis of single-particle momentum eigenstates $\{|\vec{x}\rangle\}$, Eq.~(\ref{SPstates}),
where fermionic operators act on a conventional vacuum
state.
We can easily compute the matrix elements in such a momentum base from
the mode expansions, 
Eq.~(\ref{ModeDecomp}). The relevant
operators are the contractions $\hat{P}^-_{ren}$ and the singular (Coulomb) part
of the parton-conserving interaction $\hat{P^-_{PC,s}}$
%
\beas
\langle \vec{x}|\hat{H}^-_{ren}|\vec{y}\,\rangle&=&
2\sum_{j=1}^{r=r'}\int_{0}^{y_j}\frac{dp}{(y_j-p)^2}\delta(\vec{x}-\vec{y})=
2\sum_{j=1}^{r=r'}\int_{0}^{x_j}\frac{dp}{(x_j-p)^2}\delta(\vec{x}-\vec{y}),\\
\langle \vec{x}|\hat{H}^-_{PC}|\vec{y}\,\rangle&=&
\sum_{i=1}^{r'}\sum_{j=1}^r(-1)^{(r+1)(i+j-2)}\left(\frac{1}{(x_{i}+x_{i+1})(y_{j}+y_{j+1})}-\frac{1}{(x_{i}-y_{j})^2}\right)\\
&&\qquad\qquad
\times\delta(x_i+x_{i+1}-y_j-y_{j+1})\delta(\vec{x}_{i,Sp}-\vec{y}_{j,Sp}),\\
\langle \vec{x},r'|\hat{H}^-_{PV}|\vec{y},r\rangle&=&
-\sum_{i=1}^{r'}\sum_{j=1}^{r}\delta(\vec{y}_{j,{Sp}}-\vec{x}_{i,{Sp}})\\
&&\times\left[
\left(\frac{1}{(y_{j+2}+y_{j+1})^2}-\frac{1}{(y_{j}+y_{j+1})^2}\right)
\delta^{r}_{r'+2}\delta(y_j+y_{j+1}+y_{j+2}-{x_i})\right.\\
&&\quad\,
\left.\left(\frac{1}{(x_{i+2}+x_{i+1})^2}-\frac{1}{(x_{i}+x_{i+1})^2}\right)
\delta^{r+2}_{r'}\delta(x_i+x_{i+1}+x_{i+2}-{y_j})\right],
\eeas
where $x_{Sp}, y_{Sp}$ are {\em spectator momenta}.
The matrix elements do not contain a
summation over in- or outgoing momenta. This summation/integration
appearing in the
integral equation 
is a consequence of the {\em action} of the Hamiltonian
on the states $|\Phi\rangle$, leading to the appearance of the wavefunctions
$\phi(\vec{k})=\langle \vec{k}|\Phi\rangle$. 
Note that the
non-singular term is written manifestly symmetric in in- and outgoing momenta.
Then
\beq\label{MatrixElements}
\langle \phi_{\vec{\mu}}|\hat{H}_{full}|\phi_{\vec{\nu}}\rangle
=\int_{uHS}\!\!d\vec{x}\int_{uHS}\!\!d\vec{y}\,\,\,
\langle \phi_{\vec{\mu}}|\vec{x}\rangle\langle\vec{x}|\hat{H}_{full}
|\vec{y}\rangle\langle\vec{y}|\phi_{\vec{\nu}}\rangle,
\eeq
where we {\em conventionally}
integrate over ``the'' Hilbert space, i.e.~one {\em unique Hilbert
  space} cell each.  Owing to the symmetry structure of the theory, we
can do better here in view of ensuing numerical effort if we rewrite
this integral. Paradoxically, we do better if we {\em enlarge} the
domain, because we can subsume the cyclic permutations.  In the end,
we will be able to write the eigenvalue problem of the adjoint theory
in the same {\em form} as the fundamental problem
\cite{tHooft}. Incidentally, `t Hooft used the ``trick'' of writing
the Hamiltonian as a scalar product to show that it is hermitian, not
to tame the singularity; the latter is of importance to us.


To proceed it is salutary to distinguish the Hamiltonian matrix element proportional to
\[
\sum_{ij}\frac{\delta(x_i+x_{i+1}-y_{j}-y_{j+1})\delta(\vec{x}_{rest,i}
-\vec{y}_{rest,j})}{(x_{i}-y_{j})^2}
\]
from the wavefunction part
\beq\label{DoubleDiff}
  [\phi(\vec{x})-\phi(\vec{y})]_{\mu\nu}:=\frac{1}{2}[\phi_{\mu}(\vec{x})-\phi_{\mu}(\vec{y})][\phi_{\nu}(\vec{x})-\phi_{\nu}(\vec{y})],
\eeq
which does not carry $i,j$ indices, i.e. is not part of the cyclic
permutations. Confusingly, the eigenfunctions themselves are sums over all
permutations $g\in{\cal G}$ and are affected by the
delta-function variable substitutions. The point is that the enlarging
of the integral domain affects only the matrix elements, not the wavefunctions.
To wit
\beas
&&\!\!\!\!\!\!\!\!\int_{uHS}\!\!\!\!d\vec{x}\int_{uHS}\!\!\!\!d\vec{y}
\sum_{ij}
\frac{[\phi(\vec{x})-\phi(\vec{y})]_{\mu\nu}}{(x_{i}-y_{j})^2}\Delta_{ij}
=\int_{uHS}\!\!\!\!d\vec{x}\int_{uHS}\!\!\!\!d\vec{y}
\frac{[\phi(\vec{x})-\phi(\vec{y})]_{\mu\nu}}{(x_{1}-y_{1})^2}\Delta_{11}\\
&&\!\!\!\!\!\!\!\!+\int_{uHS}\!\!\!\!d\vec{x}\int_{uHS}\!\!\!\!d\vec{y}
\frac{[\phi(\vec{x})-\phi(\vec{y})]_{\mu\nu}}{(x_{1}-y_{2})^2}\Delta_{12}
+\int_{uHS}\!\!\!\!d\vec{x}\int_{uHS}\!\!\!\!d\vec{y}
\frac{[\phi(\vec{x})-\phi(\vec{y})]_{\mu\nu}}{(x_{1}-y_{3})^2}\Delta_{13}+\ldots\\
&&\!\!\!\!\!\!\!\!=\int_{uHS}\!\!\!\!d\vec{x}\int_{uHS}\!\!\!\!d\vec{y}
\frac{[\phi(\vec{x})-\phi(\vec{y})]_{\mu\nu}}{(x_{1}-y_{1})^2}\Delta_{11}
+\int_{uHS}\!\!\!\!d\vec{x}\int_{{\cal C}uHS}\left|\frac{\partial\vec{y}}{\partial\vec{y}^{\,'}}\right|d\vec{y}^{\,'}
\frac{[\phi(\vec{x})-\phi(\vec{y}^{\,'})]_{\mu\nu}}{(x_{1}-{y}^{'}_{2})^2}\Delta'_{12}+\ldots\\
&&\!\!\!\!\!\!\!\!=\int_{cHS}\!\!\!\!d\vec{x}\int_{cHS}\!\!\!\!d\vec{y}\,\,
\frac{[\phi(\vec{x})-\phi(\vec{y})]_{\mu\nu}}{(x_{1}-y_{1})^2}\Delta_{11},
\eeas
where we 
have defined
\beas
\Delta_{ij}&:=&\delta(x_i+x_{i+1}-y_{j}-y_{j+1})\delta(\vec{x}_{rest,i}-\vec{y}_{rest,j}),\\
\Delta'_{ij}&:=&\delta(x_i+x_{i+1}-y'_{j}-y'_{j+1})\delta(\vec{x}_{rest,i}-\vec{y}^{\,'}_{rest,j}),
\eeas
and
the union of all unique Hilbert space cells connected
to the first one ($x_1\le x_{i}\,\,\forall i\neq 1$) by one of
the $r-1$ cyclic permutations
\[
\mbox{cHS}:=\mbox{uHS}\cup{\cal C}\mbox{(uHS)}\cup{\cal C}^2\mbox{(uHS)}
+\ldots\cup{\cal C}^{r-1}\mbox{(uHS)}.
\]
In the derivation, we have used $\int_{g(uHS)}d\vec{x} \,\,f(\vec{x})=\int_{uHS}d\vec{x} \,\,f(g^{-1}\vec{x})$, {\em cf.~}Eq.~(\ref{GBEtrafo}) and
$({\cal C}^{-1}\vec{y}\,')_{k+1}=y'_k$. Finally,
$\phi_r(\vec{x}^{\,'})=\phi_r({\cal C}^j\vec{x})=(-1)^{(r+1)j}\phi_r(\vec{x})$,
and the Jacobian of the
transformation induced by the cyclic permutation
\beq\label{CyclicTrafo}
x_i=x'_{i+1} \quad \forall i<r-1, \qquad x_{r-1}=1-\sum_j^{r-1} x'_j
\eeq
is
\[
\left|\frac{\partial\vec{x}}{\partial\vec{x'}}\right|=\left|\frac{\partial( x_1,x_2,\ldots,x_{r-1})}{\partial({x'_1,x'_2,\ldots,x'_{r-1}})}\right|=(-1)^{r+1}.
\]
In other words, we are performing a coordinate transformation
$\vec{y}\rightarrow \vec{y}^{\,'}={\cal C}\vec{y}$, in which the integral domain
gets mapped $uHS\rightarrow uHS'={\cal C}uHS$, and the effect of the inverse
of Eq.~(\ref{CyclicTrafo}) is to bring down the index of the momentum
fractions.
Apparently, there are two ways to interpret the integral $\int_{{\cal C}uHS}\left|\frac{\partial\vec{y}}{\partial\vec{y}^{\,'}}\right|d\vec{y}^{\,'}$: either
with rather complicated boundaries in the original variables $\vec{y}$,
or as an almost trivial copy with $y_j\rightarrow y_{j+1}$ in the new variables,
in which $y_1$ does not appear explicitly. Note that this works only with
cyclic permutations ${\cal C}^k$, under which
the Hamiltonian is explicitly symmetrized\footnote{Again, there is no quantum
  (symmetry) number $C$.}.

This means that instead of summing explicitly over in- and out-permutations,
we can simply push the integral limits to include
them, i.e. integrate over the {\em union} cHS.
As an added bonus this simplifies the integral limits, and the associated
cell volume is
\[
\int_{cHS} d\vec{x}=\int_0^1dx_1\prod_{j=2}^{r-1}\int_0^{1-\sum^{j-1}_{k=1}x_k}dx_j
=\int_0^1dx_1\int_0^{1-x_1}dx_2\int_0^{1-x_1-x_2}dx_3\cdots=\frac{1}{(r-1)!}.
\]
Recasting this in the notation of Ref.~\cite{tHooft}
($\phi_{\mu}=:\psi, \phi_{\nu}=:\varphi$),
the generalization of the 't Hooft trick,
i.e.~Eq.~(27) of \cite{tHooft}, reads
\[
\langle\psi|\hat{H}\varphi\rangle=\frac{1}{2}
\int_{cHS} \!\!d\vec{x}\int_{cHS} \!\!d\vec{y}
\,\,
\frac{[\psi^*(\vec{x})-\psi^*(\vec{y})][\varphi(\vec{x})-\varphi(\vec{y})]}{(x_1-y_1)^2}\Bigg|_{y_2=x_1+x_2-y_1\atop {y_3=x_3, y_4=x_4,...}}.
\]
This is remarkable, because it allows us to treat the much more involved adjoint
theory on the same footing as the fundamental theory.
We can now evaluate the matrix elements. We shift this technical
work to the Appendix~\ref{AppxHamMatrixElements} to 
focus on the results, i.e.~the eigensolutions of Eq.~(\ref{ABFEVP}).

\section{Results and Insights}
\label{SecResults}

Contrary to DLCQ, in eLCQ we have to labor to evaluate matrix
elements, but then the hard work is done: we have a Hamiltonian matrix
of modest dimensions (a few hundred rows and columns at most), and if we
separate its salient parts (singular, regular, mass, pair creation),
we can assemble the full Hamiltonian at will to study dependence
on parameters and importance of interaction. While this is a typical
numerical study, we can also look at the matrix elements themselves,
and get insights as to which part of the Hamiltonian
the bound state mass comes from for different states, and what the interplay
between states or role of sets of states is. 

In the latter realm are the following results. 
  The mass of the lowest state (a three-parton fermion) is entirely created
  by the parton-diagonal, regular interaction. In fact, we show
  in Appx.~\ref{AppxRegular} that the regular matrix element for all states
  with vanishing excitation numbers $\vec{\nu}=\vec{0}$ is 
  \beq\label{ZeroMass}
  \langle 0^{r-1}|\hat{P}^-_{PC,ns}|0^{r-1}\rangle=\frac{g^2 N}{\pi} r(r-1).
  \eeq
  We will see below that these states are typically insensitive
  to parton-number mixing. Modifications to the mass value in
  Eq.~(\ref{ZeroMass})
  come from mixing within the same parton sector. 
  But the only
  massless states of the (asymptotic) theory are fermionic and appear at
  alternating $T_{state}$, so in the $TIS=+++$ sectors, since all statelets
  have to have the same sign. 
  Because the three-parton state receives
  just a small correction $\Delta M^2=5.7\frac{g^2N}{\pi}$
  and there is no massless five-parton state, {\em and}
  the massless $r=7$ parton state receives a large regular correction
  of $\Delta {M}^2\approx r(r-1)\frac{g^2N}{\pi}$, the lowest (three-parton)
  state is
  basically protected against mixing due to the large mass differences
  of the states.

\begin{figure}[t]
\centerline{
\psfig{file=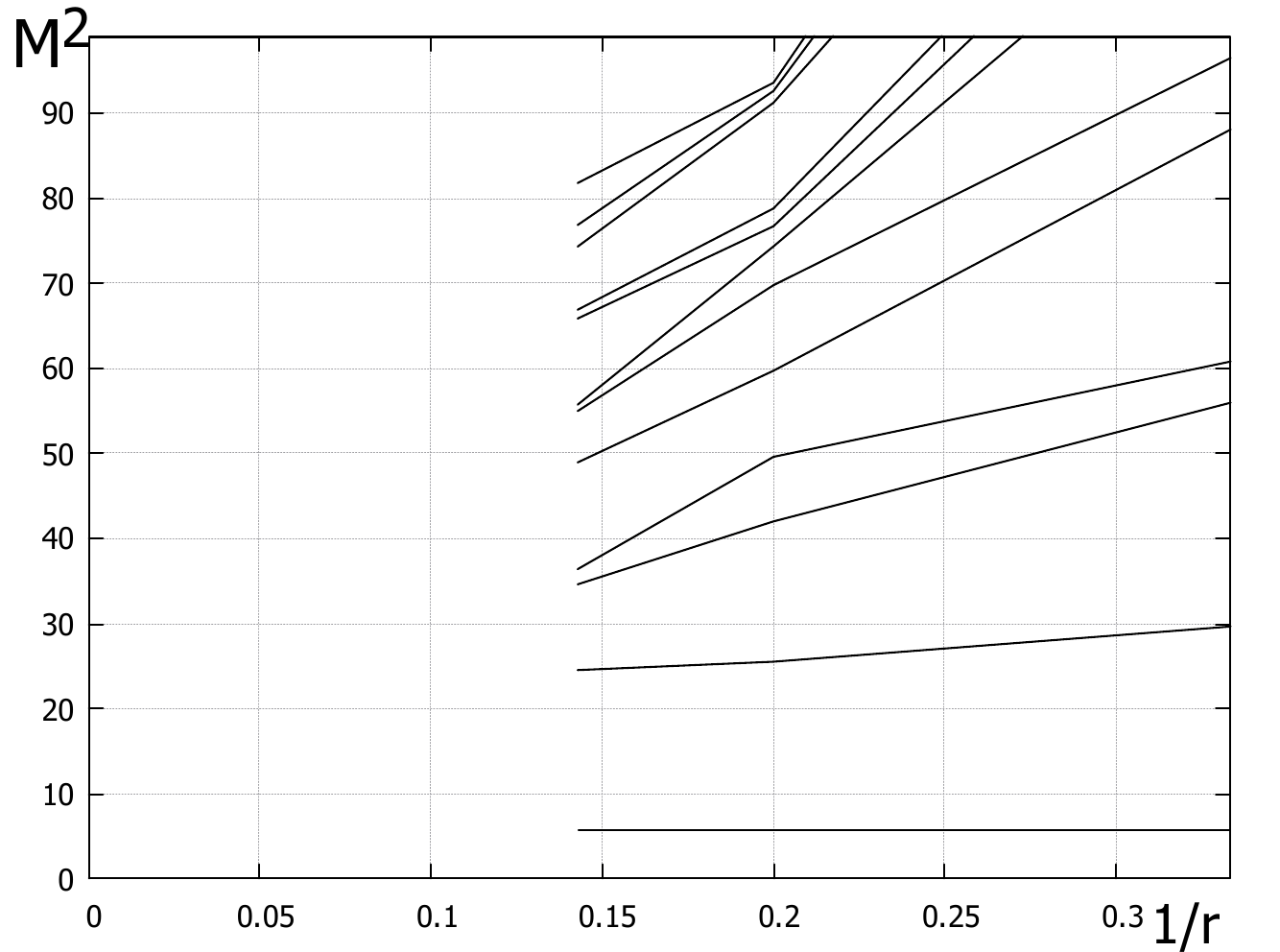,width=7.8cm}
\psfig{file=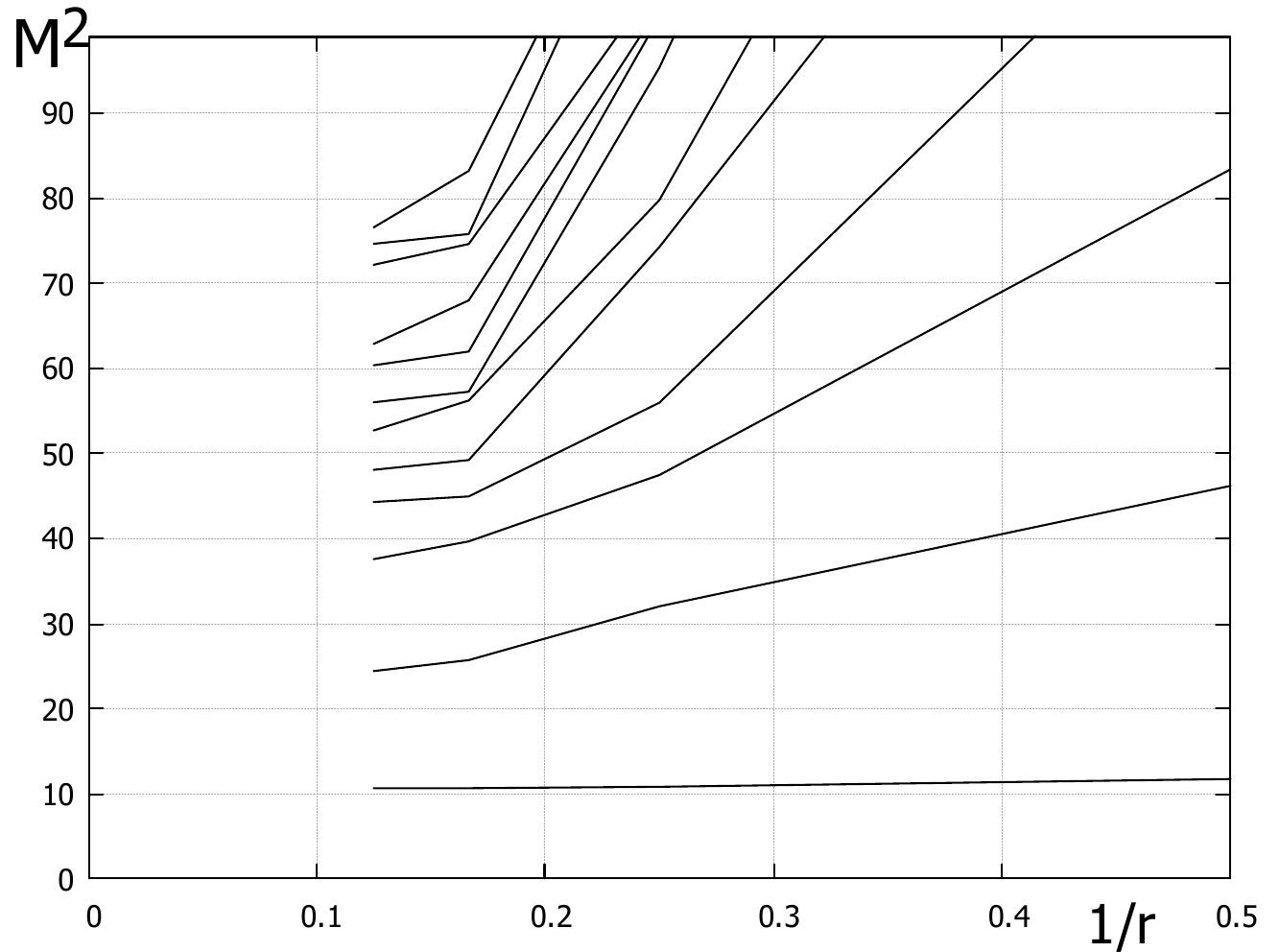,width=7.8cm}
}
\caption{Bound state masses (squared) as a function of the (maximal)
  inverse parton number
  $1/r$ in the massless fermionic $T_s\pm$ and the massless bosonic
  $T_s+$ sectors.
\label{eLCQConvergence}}
\end{figure}
%

Let's take a look at the numerical results. We should first check convergence.
As expected, convergence without pair creation is very good.
In the isolated, fixed parton-number sectors we reach percent accuracy with ten
states or less, at least for the lowest states. On the other hand,
convergence with parton number is problematic in the massless sectors of the
theory. Fig.~\ref{eLCQConvergence} shows that the ground state has well, the
first excited state somewhat converged by the time the seven(eight) parton
sector has been included in the fermionic(bosonic) sectors.
The higher states have sizable contributions from higher parton
sectors, although some of the higher bosonic masses seem pretty well converged
by $r=8$. This is what we predicted in
Sec.~\ref{GroundStates}, where we found that the lowest asymptotic fermionic
states have masses independent of parton number, whereas the masses of their
bosonic counterparts grow linearly with $r$, suppressing mixing.

Things look quite differently in the massive theory,
Fig.~\ref{FigDLCQr6mu4Tm}(a). At the supersymmetric
point $m=\frac{g^2N}{\pi}$, most masses have converged after
three parton sectors have been included. Apparently, it is energetically
expensive to create parton pairs, and the asymptotic spectrum is a good
approximation of the full solution.

\begin{figure}[ht]
\centerline{
\psfig{file=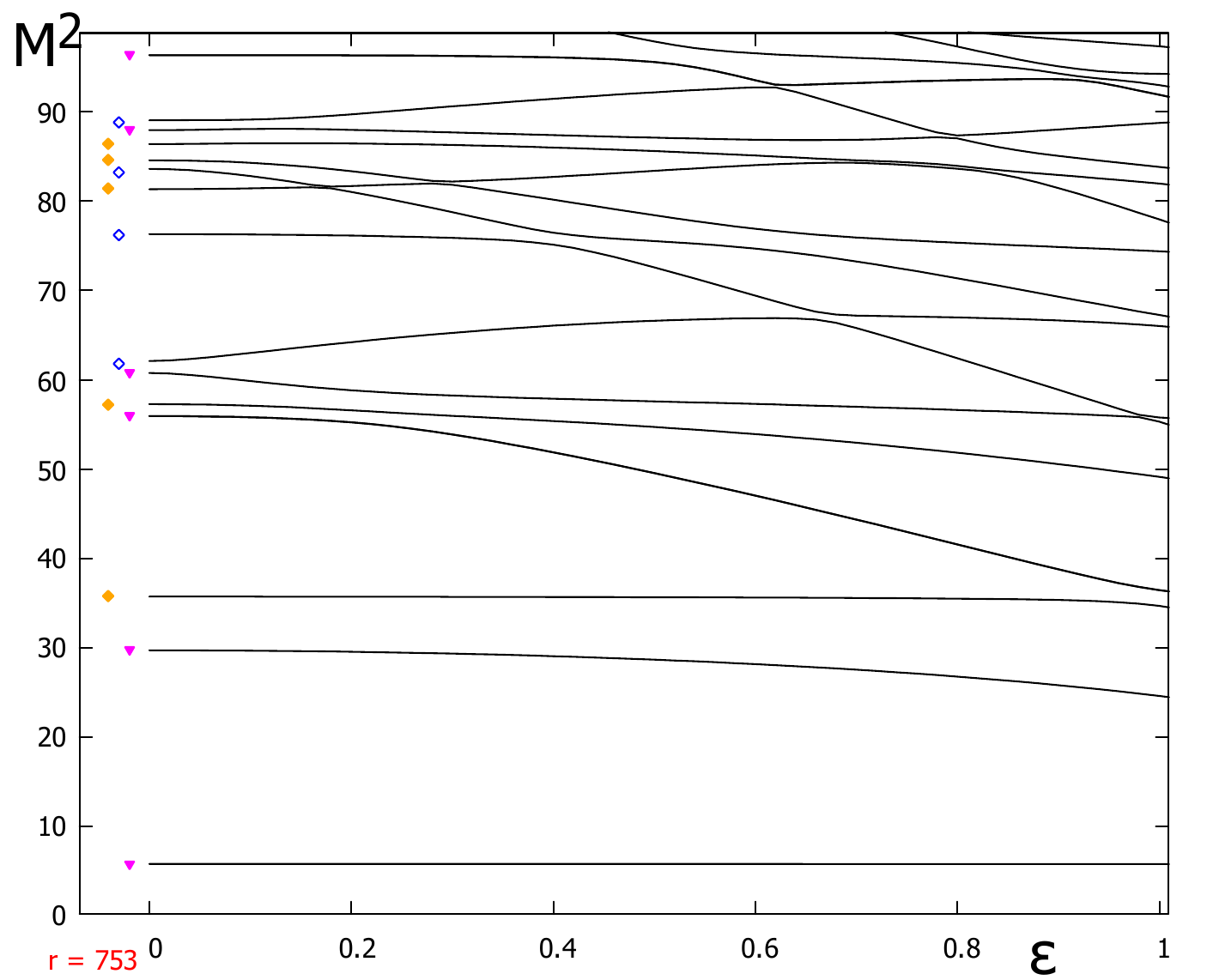,width=7.8cm}
\psfig{file=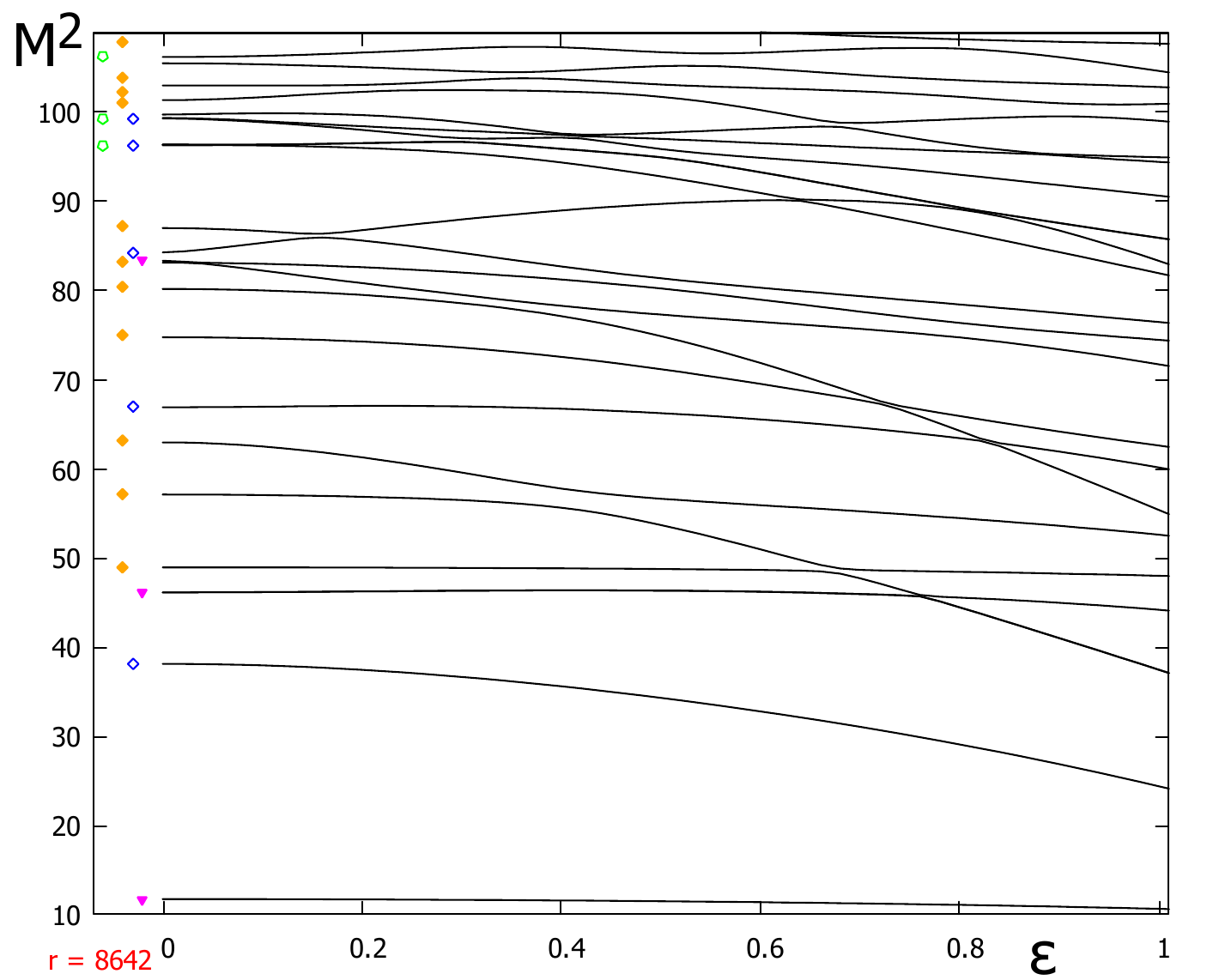,width=7.8cm}
}
\caption{Bound state masses (squared) as a function of the parton-number violation parameter $\varepsilon$
in the massless fermionic $T_s\pm$ and the massless bosonic $T_s+$ sectors; the eigenvalue trajectories
 in the massive sectors are essentially horizontal lines.
\label{eLCQvseps}}
\end{figure}
%

\subsection{The Massless Theory}

Just how important is pair creation in the massless theory?
If we plot masses as a function of the
parton-number violation parameter $\varepsilon$, where the
asymptotic(full) theory
has $\varepsilon=0(1)$, Fig.~\ref{eLCQvseps},
we see that some states' masses are
quite insensitive, while others depend substantially on $\varepsilon$. In fact,
there seems to be a cascade of $\varepsilon$-dependent states. 
These states are impure in parton number, and intriguingly
some of them have been identified
previously \cite{GHK,Kleb21} as threshold bound states, i.e. multi-particle
states. For instance, the second lowest bosonic state in Fig.~\ref{eLCQvseps},
which
starts out (as depicted in the left portion of the graph) as a four-parton
state, has a 19\%, 57\%, 17\%, 2\% probability to be a 2, 4, 6, 8 parton state.
The pattern continues at higher masses. 
It seems thus that there are two types of states.
It is not clear what mechanism is at work to protect the pure states by heavily
mixing the impure states. It could be that the impure states decouple from the
rest of the spectrum in the large parton limit or, in DLCQ,
in the continuum limit. It would be interesting to apply eLCQ to the
bosonized theory to see whether the approximate multi-particle states
decouple (as in: they are present in \cite{UT1} only because the
continuum limit has not been taken), or whether they
are a genuine part of the theory. 

Judging from the results at hand, 
eLCQ has limited capabilities to contribute positively
to the debate about the massless theory. That said,
it does
provide evidence that the massless theory has a very different spectrum because
its asymptotic eigenstates are in crucial aspects different from their massive
counterparts.
As we will see later, eLCQ also points to a problem of other approaches with the
massless theory: due to its singularities it becomes near impossible to produce
accurate, quantitative results with a simple, rigid IR regulator such as DLCQ.

\subsection{The massive theory and the supersymmetric point}

The massive theory's spectrum is quite different from its massless counterpart
which may shed some light on the controversy as to whether the
massless theory is screening or confining \cite{Kleb21, Jacobson}.
Note that we argue here from a basis function point of view. This may
seem naive, but keep in mind that it is the physics
(e.g. representation of fields) that determines the ``boundary
conditions'' and therefore the appropriate set of basis states.

In the massive regime, all excitation numbers are
even integers; the odd excitation numbers of the bosonic massless
states would make it impossible to match fermionic and bosonic mass
eigenvalues.  All massive eigenfunctions are built form the same trig
function, whereas in the massless sectors they are opposite (sine goes
with cosine in the adjacent parton sector). Without this
feature (following automatically from the symmetry structure of the
theory) a supersymmetric point at $m=\frac{g^2N}{\pi}$ would be impossible.

\begin{figure}[ht]
\centerline{
\psfig{file=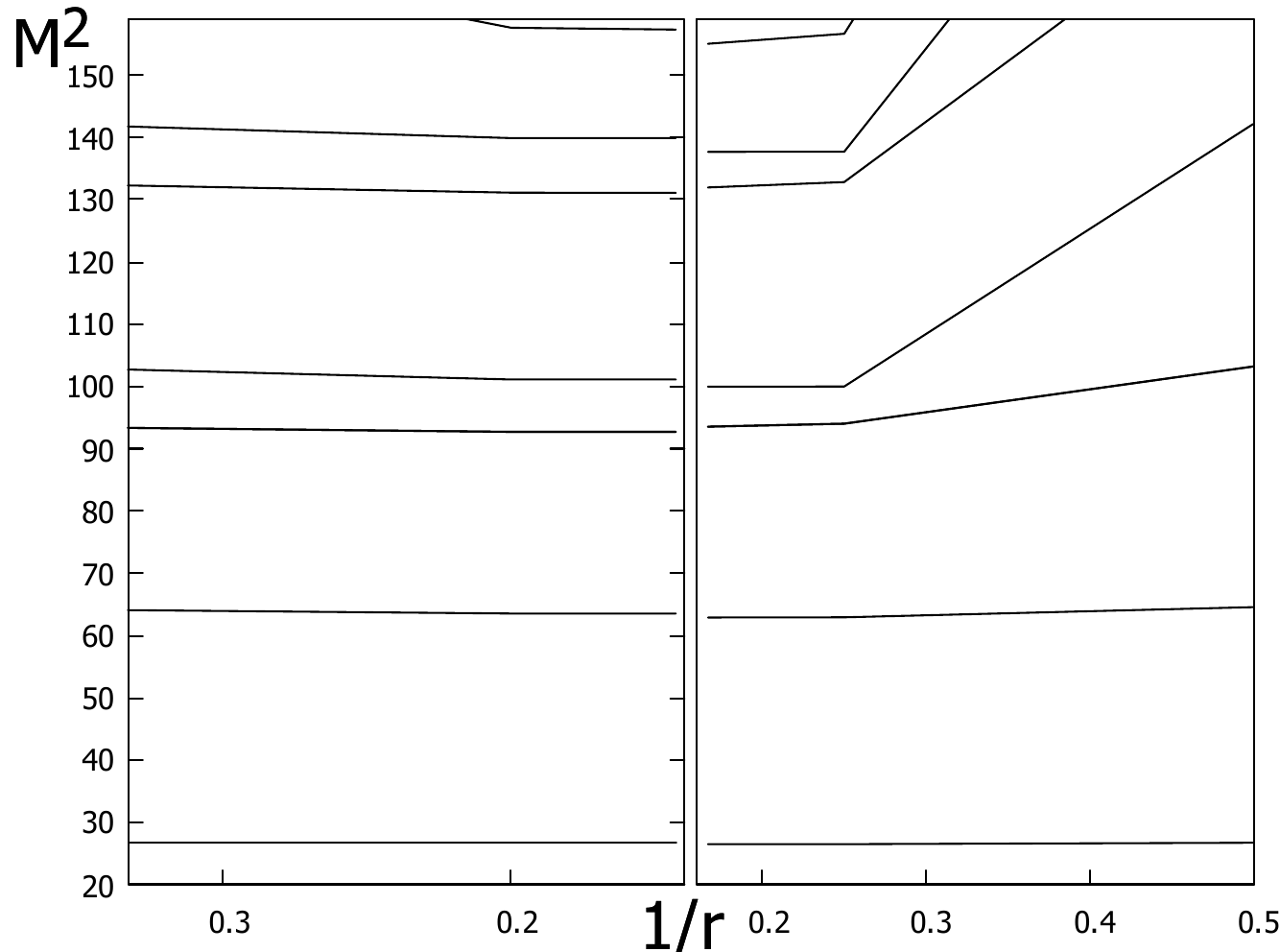,width=7.8cm}
\psfig{file=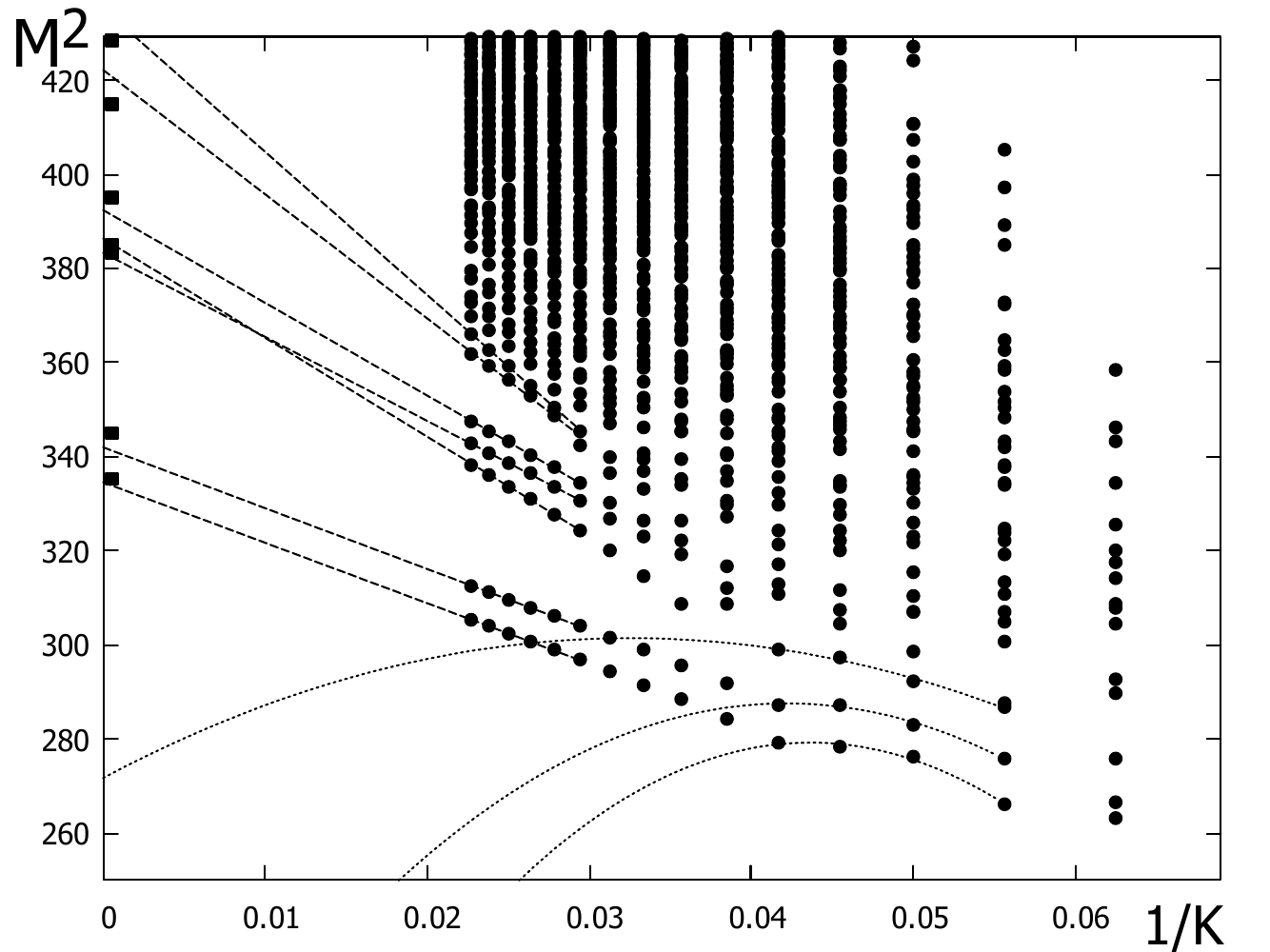,width=7.8cm}
}
\caption{(a) Convergence and comparison of fermionic (left) and bosonic (right) eigenvalues at the
  supersymmetric point ($m=\frac{g^2N}{\pi}$ or $\mu=1$)
  as a function of inverse parton number $1/r$. Discrepancy of corresponding eigenvalues
is at most 1.5\%.
(b) Comparison of DLCQ and eLCQ bound state masses (squared) in the  
asymptotic massive six parton sector with $T+$ and $\mu=4$.
DLCQ results are dots plotted as a function of inverse harmonic resolution $1/K$; eLCQ results
appear as squares at infinite resolution. 
Linear fits (dashed lines) to the DLCQ masses at highest $K$ agree well with eLCQ.
Quadratic fits to data up to $K=24$ lead to unreasonable results, even though the behavior of
the lowest three mass trajectories appears consistent.
\label{FigDLCQr6mu4Tm}}
\end{figure}

The massive sectors have the least symmetric states,
i.e.~most disjunct statelets, which makes computing matrix elements
expensive. On the positive side, the asymptotic eigenstates are
very good approximations to the full eigenstates; the diagonal
Hamiltonian matrix blocks (singular, regular, mass terms) are
dominated by their diagonal elements, whereas the pair-creation matrix
elements are small. Hence, there is very little coupling between
sectors of different parton number (observe the flatness of the
eigenvalue trajectories in Fig.~\ref{FigDLCQr6mu4Tm}(a)), but also
states with the same parton number hardly mix. Note that this feature
is mostly independent of the mass of the fermions, as the singular and
regular blocks just depend on the symmetry quantum numbers $TIS$, not
on mass. In other words, there is a {\em non-continuous} difference
between the massive and massless theory, favoring a different
(screening) behavior of the latter.

Can we understand this qualitatively? After all, we are saying that a
two-parton bosonic state yields the same (diagonal) element as a
three-parton fermionic state. We consider the lowest states as the
simplest case without loss of generality, since we need to have
one-to-one matching of quasi-isolated states.  From the singular, regular
and mass term contributions we gather
$23.2+0+6.2\mu=29.8$ as a crude estimate for the lowest two-parton
mass in units $g^2N/\pi$ (true: 26.7),
whereas for the three-parton state we have
$14.4+3.7+11.3\mu=29.4$, so this agrees pretty well.
The mechanism behind this ``adjustment'' of singular, regular and mass
contributions is not obvious. Consider that 
in order to obtain the matrix elements, we integrate a
single-variable sine over a one-dimensional domain vs.~a $\sin\pi(n_1
x_1+n_2 x_2)$ over a much more complicated
domain\footnote{Technically, it is the product of two such sines, but
  we use trig identities.}. 
Of course, supersymmetry guarantees this degeneracy of fermion and boson
masses,
as shown in Ref.~\cite{Kutasov94} by using the supersymmetric
generator\footnote{Note that $G_0$ acts on operators not the
  wavefunction, so we need the same trig function in supersymmetric
  partner sectors.}  $G_0$. Nonetheless, it is amusing to watch this unfold
numerically --- even though it is the equivalent of showing (by group theory)
that the product two rotations around two different axes is always a rotation.

Overall, eLCQ works well for the massive theory. To a fair approximation the
 asymptotic states describe the solutions of the full theory, and the
 supersymmetric degeneracy of boson and fermion bound-state masses is
 reproduced.

%
%

\section{Discussion and Conclusion}
\label{SecDiscussion}

In this work we presented a complementary calculation of the spectrum
of two-dimensio\-al adjoint QCD. By using a basis-function approach
based on the asymptotic spectrum of the theory generated via the eLCQ
algorithm, we work in the (momentum) continuum limit. The {\em Achilles heel}
of the method is the relatively small number of parton sectors
in which the Hamiltonian matrix elements can be calculated --- at least
by brute force
methods. Leveraging the insights gained in the construction of the complete
asymptotic eigenfunction spectrum, we understand that this is not
a problem in the massive theory $m>0$,
where pair creation is unimportant.
In the {\em massless} theory we find states whose asymptotic masses are
independent of parton number. We therefore expect the generic eigenstate
in this sector to have substantial contributions from all parton sectors.
As first noticed in \cite{DalleyKlebanov}, this is not the case for the
lowest states; they are pure in parton number. This can be understood via eLCQ
from the severe constraints of possible excitation numbers allowed by the
symmetries of the theory. These symmetries explain the properties
of the lowest states in remarkable detail, allowing for a good estimate of
the masses, even though the asymptotic approximation assumes {\em high}
excitation numbers.

Overall, our results are in fair agreement with previous DLCQ-based
work \cite{GHK,Kleb21}.
This shows that discretization or compactification of the theory yields
qualitatively correct results, even if accuracy of bound-state masses
becomes an issue. Note that in DLCQ(eLCQ) mass trajectories increase(decrease)
as one approaches the continuum(infinite parton number) limit.
In this light, it is concerning that there is a
discrepancy even in the ground state mass
\[
M^2_{F0}=5.72|_{DLCQ}\frac{g^2N}{\pi}\ge 5.69|_{eLCQ}\frac{g^2N}{\pi}.
\]

After completing this work, we appreciate the great advantage of DLCQ
to consistently (if coarsely at available $K$) approximating 
all states, which results in a faithful representation of the underlying algebraic structure. This is paramount for representation theory
analyses \cite{Kleb21,Kleb22}, and barring any improvements on the eLCQ
approach, might be more important than accurately describing the eigenstates
at low parton number. If the advantages of the two methods could
be combined, one would have a powerful tool to analyze low-dimensional
field theories! 

In the meantime, we should point out the pitfalls in both approaches.
Take a look at Fig.~\ref{FigDLCQr6mu4Tm} depicting
the masses of six-parton states when one neglects pair creation but includes
all other interactions. Note that
this is in the massive theory, so the former is
actually a good approximation even in eLCQ. In Fig.~\ref{FigDLCQr6mu4Tm}
the DLCQ continuum limit cannot be taken
if one has data only for $K<26$. The lowest states do not
split off until a crucial resolution is reached; for higher states this
resolution is obviously higher. In fact, at $r=7$, the condition
has worsened to the point that this split (and hence the possibility of
extrapolation) does not happen until one is forced to use sparse matrix methods
(roughly 10,000 states at $K\ge 41$). Note that these large resolutions are
not possible when all parton sectors are included.
Fig.~\ref{FigDLCQr6mu4Tm} also shows that eLCQ does get it right. Its
limitation is that it essentially stops working at nine partons due to the
exponentially growing numbers of terms in the Hamiltonian and statelets.

As mentioned, the conclusions of \cite{Kleb21,Kleb22} do not depend
on the precise masses, but rather on the consistent, qualitative features
of the spectrum and the relations (degeneracies) across different sectors
of the theory.
For eLCQ one might think of going to higher parton sectors by
stochastically sampling statelets and terms.

Let us briefly discuss potential problems of discretized approaches
at finite resolution --- which are surprisingly benign as far as we
can tell from comparing eLCQ and DLCQ results.
To expose problems, we consider a hybrid method (``hLCQ''),
where we take a DLCQ
Hamiltonian matrix and sandwich it between eLCQ continuous asymptotic
eigenfunctions. One then finds that there is a crucial discrepancy
between the matrix elements computed in the two schemes.
The hybrid method fools us into diagnosing
a linear convergence in $1/K$ of the matrix elements 
towards a much smaller value than the (correct) one obtained analytically via
the eLCQ algorithm. We can trace the dominant
contributions to the matrix elements to
integrals of the form $\int^1_0 \cos(\pi(n'-n_3) x)
{\rm Si}(\pi n_1 x) dx$ in which one of the incoming excitation numbers $n_i$
is equal to the outgoing $n'$. These expressions result from similar integrals
in the function $I_{PNV}(x_1,y_1)$, see appendix, which
is diverging whenever $y_1=0$, while its integral over $y_1$ has only isolated
divergences that will eventually cancel. It is clear, that any {\em numerical}
integration --- let alone a multi-dimensional one --- will
have poor outcomes.
It is surprising that these issues with matrix elements in DLCQ
do not have more severe consequences. It must be that there is a partial
cancellation of errors which likely stems from the highest virtue of DLCQ:
consistent approximations in all parton sectors.

In conclusion, the application of the eLCQ algorithm to QCD$_{2A}$
generates a continuous (at fixed parton number) method to compute the spectrum
of the theory. It is complementary to discretized approaches,
and validates certain findings within these frameworks.
While working in the continuum allows us to address some issues
with using the light-cone gauge in a discretized formulation \cite{Jacobson},
problems connected to the Hamiltonian itself are beyond the scope of the
present work.
Since eLCQ is limited
to small parton numbers, it would be interesting to apply it to the 
bosonized version of QCD$_{2A}$,
where the effective parton number is essentially halved.
It is not clear if this is feasible due to the symmetries being
compromised by the Kac-Moody commutator.
In general, however, one should be able with the present approach
to figure out more easily which (formulations of) theories result
in eigenstates pure in parton number, and for which pair production is
crucial.

\section*{Acknowledgments}

Discussions with D.G.~Robertson
about his work in \cite{CartorRobPin97} are gratefully acknowledged, as is the
hospitality of the Ohio State University's Physics 
Department, where most of this work was completed.
I thank Otterbein University for providing a work environment in
which projects such as the present can be tackled. 

\begin{appendix}


\section{Evaluating Matrix Elements I --- General Method}
\label{AppxHamMatrixElements}

We now apply the results of Sec.~\ref{SecNumericalSolution}, i.e.~the enlarged
integral domain, to
evaluating the matrix elements for arbitrary parton number $r$ with
the {\em spectator variables} $\vec{x}_{Sp}$ and $\vec{y}_{Sp}$.
We start with
\bea\label{GenrStart}
\langle\vec{\mu}_r|\hat{H}_{0}|{\vec{\nu}_r}\rangle\!\!\!&=&\!\!\!
\int_{cHS} \!\!d\vec{x}\int_{cHS} \!\!d\vec{y}
\,\,\delta(x_1+x_2-y_1-y_2)\delta^{r-3}(\vec{x}_{Sp}-\vec{y}_{Sp})
\frac{[\phi(\vec{x})-\phi(\vec{y})]_{\mu\nu}}{(x_1-y_1)^2}\nonumber\\
\!\!\!&=&\!\!\!
\int_0^1 \!\!\!dx_1\int_0^{1-x_1} \!\!\!\!dx_2\int d^{r-3}\vec{x}_{Sp} \int_0^{x_1+x_2}\!\!\!
dy_1\frac{[\phi(\vec{x})-\phi(\vec{y})]_{\mu\nu}}{(x_1-y_1)^2}\Big|_{y_2=x_1+x_2-y_1\atop {\!\!\!\!\!\!\!\!\!\!\!\!\!\! \vec{y}_{Sp}=\vec{x}_{Sp}}},
\eea
where the upper limit on the last integral follows from $y_2\ge 0$.
To split off the non-interacting partons (spectators),
we have to commit to the actual form of our eigenfunctions,
cf.~Eq.~(\ref{Normalization})
\[
\phi_{I+,\nu}={\cal N}\sum_{g\in{\cal G}} \cos \left(\pi g \vec{\nu}\!\cdot\!\vec{x}\right)
  \qquad\mbox{ or }\qquad
\phi_{I-,\nu}={\cal N}\sum_{g\in{\cal G}} \sin \left(\pi g \vec{\nu}\!\cdot\!\vec{x}\right).
\]
To simplify notation, call $(g\nu)_i\equiv n_i$, and use the sum-to-product
trig identities to split off the {spectator variables} $\vec{x}_{Sp}$ and
$\vec{y}_{Sp}$ in the {\em double difference}
$[\phi(\vec{x})-\phi(\vec{y})]_{\mu\nu}$,
Eq.~(\ref{DoubleDiff}).
This leads to
\beas
[\phi(\vec{x})-\phi(\vec{y})]^{+}_{\mu\nu}&=&\frac{1}{2}\left[\left\{\cos\pi(m_1x_1+m_2x_2)-\cos\pi(m_1y_1+m_2y_2)\right\}
  \cos\pi \vec{m}\!\cdot\!\vec{x}_{Sp}\right.\\
  &&\quad-\left.\left\{\sin\pi(m_1x_1+m_2x_2)-\sin\pi(m_1y_1+m_2y_2)\right\}
    \sin\pi \vec{m}\!\cdot\!\vec{x}_{Sp}\right]\\
 &&\times \left[\left\{\cos\pi(n_1x_1+n_2x_2)-\cos\pi(n_1y_1+n_2y_2)\right\}
  \cos\pi \vec{n}\!\cdot\!\vec{x}_{Sp}\right.\\
  &&\quad-\left.\left\{\sin\pi(n_1x_1+n_2x_2)-\sin\pi(n_1y_1+n_2y_2)\right\}
    \sin\pi \vec{n}\!\cdot\!\vec{x}_{Sp}\right],
\eeas
and similar for $\phi_{I-}$,
where we set $\vec{x}_{Sp}=\vec{y}_{Sp}$ due to the second delta function
in Eq.~(\ref{GenrStart}).
So we get four terms for $I+$ (and similar for $I-$)
{
\beas
[\phi(\vec{x})-\phi(\vec{y})]^{+}_{\mu\nu}&\!\!\!\!=\!\!\!\!&D^{00}_{\bar{\mu}\bar{\nu}}\cos\pi \vec{m}\!\cdot\!\vec{x}_{Sp}\cos\pi \vec{n}\!\cdot\!\vec{x}_{Sp}
- D^{01}_{\bar{\mu}\bar{\nu}}\cos\pi \vec{m}\!\cdot\!\vec{x}_{Sp}\sin\pi \vec{n}\!\cdot\!\vec{x}_{Sp}\\
&&\!\!\!\!- D^{10}_{\bar{\mu}\bar{\nu}}\sin\pi \vec{m}\!\cdot\!\vec{x}_{Sp}\cos\pi \vec{n}\!\cdot\!\vec{x}_{Sp}
+ D^{11}_{\bar{\mu}\bar{\nu}}\sin\pi \vec{m}\!\cdot\!\vec{x}_{Sp}\sin\pi \vec{n}\!\cdot\!\vec{x}_{Sp},
\eeas
}
where the {double differences} are defined as 
\bea\label{DD}
D^{00}_{\bar{\mu}\bar{\nu}}(x_1,x_2,y_1)&:=&\frac{1}{2}
\left\{\cos\pi(m_1x_1+m_2x_2)-\cos\pi(m_1y_1+m_2y_2)\right\}\nonumber\\
&&\times\left\{\cos\pi(n_1x_1+n_2x_2)-\cos\pi(n_1y_1+n_2y_2)\right\},\nonumber\\
D^{01}_{\bar{\mu}\bar{\nu}}(x_1,x_2,y_1)&\!\!\!:=\!\!\!&\frac{1}{2}
\left\{\cos\pi(m_1x_1+m_2x_2)-\cos\pi(m_1y_1+m_2y_2)\right\}\nonumber\\
&&\times\left\{\sin\pi(n_1x_1+n_2x_2)-\sin\pi(n_1y_1+n_2y_2)\right\},
\eea
etc., and $\bar{\mu}\bar{\nu}$ is short for $m_1,m_2,n_1,n_2$.
Recall $y_2=x_1+x_2-y_1$.
\noindent Analytically integrating over $\vec{x}_{Sp}$ is
straightforward if tedious
for large $r$. We label the results as follows
\bea\label{SpectatorIntegrals}
F^{ab-}_{mn}(x_1,x_2)&:=&\int d^{r-3}\vec{x}_{Sp}
\sin_a\left(\pi\vec{m}\!\cdot\!\vec{x}_{Sp}\right)
\sin_b\left(\pi\vec{n}\!\cdot\!\vec{x}_{Sp}\right),
\eea
where $a,b\in\{0,1\}$,  $\sin_0=\cos$, and  $\sin_1=\cos$. 
Note that $F^{01-}_{mn}=F^{10-}_{nm}$. Letting
$F^{ab+}_{mn}(x_1,x_2)=(-1)^{a+b}F^{1-a,1-b-}_{mn}(x_1,x_2)$,
we are left with the three-dimensional integral
\beq\label{GenrMatrixElement}
\langle\vec{\mu}_r|\hat{H}_{0}|{\vec{\nu}_r}\rangle^{\pm}=
\sum_{g_{in},g_{out}\in{\cal G}}\sum_{a,b=0}^1
\int_0^1 dx_1\int_0^{1-x_1} \!\!\!\!dx_2
F^{1-a,1-b\pm}_{(\vec{m},\vec{n})_{Sp}}(x_1,x_2)
\int_0^{x_1+x_2} \!\!\!\!dy\frac{D^{ab}_{\bar{\mu}\bar{\nu}}(x_1,x_2,y)}{(x_1-y)^2}.
\eeq


To evaluate the matrix element, we first integrate over the momentum $y$
exchanged between the interacting partons for various trig function combinations
\[
\tilde{I}^{+,-,0}_y(x_1,x_2):=\int_0^{x_1+x_2}dy
\frac{D^{00,11,01}_{\bar{\mu}\bar{\nu}}(x_1,x_2,y)}{(x_1-y)^2}
\]
To properly treat the singularities, the integrals are to be taken with
the prescription
\[
\int_0^1 dx_1\int_0^{1-x_1}dx_2\int_0^{x_1+x_2} \!\!\!dy:=\lim_{\epsilon\rightarrow 0}
\int_0^1 dx_1\int_0^{1-x_1} dx_2 \left( \int_0^{x_1-\epsilon} \!\!\!dy+
\int^{x_1+x_2}_{x_1+\epsilon} \!\!\!dy \right).
\]
In total we have in the $I+$ sector with $m_{\pm}:=m_1-m_2\pm(n_1-n_2)$
{\small
\bea
\tilde{I}^+_y
\!\!\!\!&=&\!\!\!-\frac{1}{x_2}[\cos \pi m_1 (x_1+x_2) \cos \pi n_1 (x_1+x_2)
	+\cos \pi (m_1 x_1+m_2 x_2) \cos \pi (n_1 x_1+n_2 x_2)\nonumber\\
	&&\qquad -\cos \pi (m_1 x_1+m_2 x_2) \cos \pi n_1 (x_1+x_2)
	 -\cos \pi (n_1 x_1+n_2 x_2) \cos \pi m_1 (x_1+x_2)]\nonumber\\
&&\!\!\!-\frac{1}{x_1}[\cos \pi m_2 (x_1+x_2) \cos \pi n_2 (x_1+x_2)
      +\cos \pi (m_1 x_1+m_2 x_2) \cos \pi (n_1 x_1+n_2 x_2)\nonumber\\
  &&\qquad    -\cos \pi (m_1 x_1+m_2 x_2) \cos \pi n_2 (x_1+x_2)
      -\cos \pi (n_1 x_1+n_2 x_2) \cos \pi m_2 (x_1+x_2)]\nonumber\\
&&\!\!\!-\frac{\pi}{2} \Big\{ m_+ \sin \pi ((m_1+n_1) x_1+(m_2+n_2) x_2)
		   [{\rm Ci\,} \pi |m_+| x_2-{\rm Ci\,} \pi |m_+| x_1]\nonumber\\
		&&\!\!\!\qquad  + m_- \sin \pi ((m_1-n_1) x_1+(m_2-n_2) x_2)
		  [{\rm Ci\,} \pi |m_-| x_2-{\rm Ci\,} \pi |m_-| x_1]\Big\}
                  \nonumber\\
  &&\!\!\!- \pi(n_1-n_2) \cos \pi (m_1 x_1+m_2 x_2) \sin \pi (n_1 x_1+n_2 x_2)
     [{\rm Ci\,} \pi |n_1-n_2| x_2-{\rm Ci\,} \pi |n_1-n_2| x_1]\nonumber\\
   &&\!\!\! -\pi (m_1\!-\!m_2) \cos \pi (n_1 x_1\!+\!n_2 x_2) \sin \pi (m_1 x_1\!+\!m_2 x_2)
     [{\rm Ci\,} \pi |m_1\!-\!m_2| x_2-{\rm Ci\,} \pi |m_1\!-\!m_2| x_1]\nonumber\\
     &&\!\!\! -\frac{\pi}{2} \Big\{\pi m_+ \cos \pi ((m_1+n_1) x_1+(m_2+n_2) x_2)
	 [{\rm Si\,}\pi m_+ x_2+{\rm Si\,} \pi m_+ x_1)]\nonumber\\
&&\qquad	  + m_- \cos \pi ((m_1-n_1) x_1+(m_2-n_2) x_2)
		   [{\rm Si\,} \pi m_- x_2+{\rm Si\,} \pi m_- x_1]\Big\}\nonumber\\
   &&\!\!\!  -\pi(n_1-n_2) \cos \pi (m_1 x_1+m_2 x_2)\cos \pi (n_1 x_1+n_2 x_2)
       [{\rm Si\,} \pi (n_1-n_2) x_2+{\rm Si\,} \pi (n_1-n_2) x_1]\nonumber\\
  &&\!\!\!       
     - \pi(m_1\!-\!m_2)  \cos \pi (n_1 x_1\!+\!n_2 x_2)\cos \pi (m_1 x_1\!+\!m_2 x_2)
       [{\rm Si\,} \pi (m_1\!-\!m_2) x_2+{\rm Si\,} \pi (m_1\!-\!m_2) x_1]\nonumber
\eea
}
and similar for $\tilde{I}^-(x_1,x_2)$  and $\tilde{I}^0(x_1,x_2)$.
We then assemble the full matrix elements by combining the
$\tilde{I}_y(x_1,x_2)$ functions with the spectator functions
$F^{ab}(x_1,x_2)$. The result are functions of $x_1,x_2$ which involve
powers, trig functions and logarithms.

Before we categorize and integrate those in the next section, we point out
that Eq.~(\ref{GenrStart}) contains a redundancy which will allow us to save at
least a factor of two in numerical effort. Namely, due to the fact that
$x_r$ does not explicitly appear yet is present in the formalism (we went to
great lengths in Sec.~\ref{SecNumericalSolution}
to keep it out of calculations),
we can replace the integral over $x_{r-1}$ with an integral over $x_r$
(same limits). This means that sandwiching the Hamiltonian between statelets
that are cyclically rotated (on both in and out sides) such that $x_{r-1}$
and $x_r$ are permuted will yield the same result.
For instance,
\[
\langle \mu;0| \hat{H}|\nu; r-1\rangle=
\langle \mu;0| \hat{H}{\cal C}^{r-1}|\nu; 0\rangle= 
\langle \mu;0| {\cal C}^{r-1}\hat{H}|\nu; 0\rangle= 
\langle \mu;r-1| \hat{H}|\nu; 0\rangle,
\]
where $|\nu; j\rangle$ is the $j$th statelet of the state characterized by the
excitation tuplet $\nu$. 

\section{Evaluating Matrix Elements II --- Integrals}

All matrix elements require us to integrate expressions of the form
\[
\frac{\sos(\pi\vec{m}\!\cdot\!\vec{x})\,\cin(\pi\vec{n}\!\cdot\!\vec{y})}
     {(x_1\pm y_1)^2}
\]
over various domains of different dimensionality\footnote{For instance,
  in the parton-number
  violating interaction, a one-dimensional $x$-integral is
  paired with a three-dimensional $y$-integral.}.
All singularities are integrable if the integrals are carefully regulated and
one evaluates only matrix elements between {\em bona fide} asymptotic states.
In particular, matrix elements between {\em statelets} might still be
singular, typically like $\ln\epsilon$ or $\ln\bar{\epsilon}$,
where the former
is a {\em spatial}, the latter a {\em excitation number} regulator\footnote{For
  example, we write ${\rm Ci}(\pi k x)$ as ${\rm Ci}(\pi k \epsilon)$ for
  $x\rightarrow 0$ and as ${\rm Ci}(\pi \bar{\epsilon}x)$ for $k\rightarrow 0$.}. 
The four fundamental categories of integrals we encounter are 
\[
\int dx\, x^n \sin_s(\pi k x),
\int dx\, x^n {\rm Si}_S(\pi q x),
\int dx\, x^n \sin_s(\pi k x)\ln x,
\int dx\, x^n \sin_s(\pi k x){\rm Si}_S(\pi q x),
\]
where $s=0(1)$ denotes a $\cos(\sin)$ and $S=0(1)$ is a ${\rm Ci}({\rm Si})$.
All matrix elements are linear combinations of integrals of these types,
e.g. part of the ``singular'' matrix element 
\beq\label{Integral9}
\int^1_0 dx_1\, x_1^{n_1}\int^{1-x_1}_0 dx_2\, x_2^{n_2}\sin_s(\pi (k_1 x_1+k_2 x_2)){\rm Si}_S(\pi q x_2).
\eeq
This makes sense, since integrating a double pole will lead to a single pole
or a sine or cosine integral, while integrating again will turn the single
pole into a logarithm (worst case). The powers of $x$ are generated by {\em
  spectator integrals} in the higher parton sectors if two excitation numbers
are equal or opposite.
We sketch the specific
evaluations for singular, regular and parton-number violating interactions
in the following subsections.

\subsection{Fundamental Integrals}
\label{FundIntegrals}

We need the following results
\bea\label{xnCinIntegrals}
\int x^n \sin(\pi(a+kx))dx
&=&\left(\sum_{j=1}^{n/2+1}\frac{(-1)^j}{(\pi k)^{2j-1}}
\frac{n!}{(n-2j+2)!} x^{n-2j+2}
\right)\cos(\pi(a+kx))\nonumber\\
&&+\left(\sum_{j=1}^{(n+1)/2}\frac{(-1)^{j+1}}{(\pi k)^{2j}}
\frac{n!}{(n-2j+1)!}x^{n-2j+1}
\right)\sin(\pi(a+kx))\nonumber\\
&=:& A_n(x)\cos(\pi(a+kx))+B_n(x)\sin(\pi(a+kx))\\
\int x^n \cos(\pi(a+kx))dx
&=& -A_n(x)\sin(\pi(a+kx))+B_n(x)\cos(\pi(a+kx)),
\eea
where $A_n=:\sum_{j=1}^{n/2+1}a_{nj}x^{n-2j+2}$, etc..
These expressions have to be evaluated at the limits
($L:=1-\sum_{i=1}^t x_i,0$) so that
\beas
\int_0^L x^n \sin(\pi(a+kx))dx&=&A_n(L)\cos(\pi(a+kL))-a_{n,n/2+1}(0)\cos(\pi a)\\
&&+B_n(L)\sin(\pi(a+kL))-b_{n,(n+1)/2}(0)\sin(\pi a),
\eeas
where
\[
a_{n,n/2+1}(0)=\frac{n!(-1)^{n/2+1}}{(\pi k)^{n+1}}, \qquad
b_{n,\frac{n+1}{2}}(0)=-\frac{n!(-1)^{\frac{n+1}{2}}}{(\pi k)^{n+1}}.
\]
Note that at the lower limit there is only at most one (constant) term,
since $j$ has to be such that the exponent of $x$ is zero.

For the second type of integrals we have ($\tilde{x}_1:=1-x_1$)
{\small
\beas
\int_0^{\tilde{x}_1} x^n {\rm Si}(\pi qx)dx\!\!\!&=&\!\!\!\frac{1}{n+1}\Big[
\tilde{x}_1^{n+1}{\rm Si}(\pi q \tilde{x}_1)+
\frac{\tilde{x}_1^{n}}{\pi q}\Big(\cos(\pi q\tilde{x}_1)-\delta^{n}_0\Big)
-\frac{n\bar{\delta}^{n}_0}{\pi q}
\int^{\tilde{x}_1}_0\!\!\!\! x^{n-1}\cos(\pi q x)dx\Big]\\
\int_0^{\tilde{x}_1} x^n {\rm Ci}(\pi qx)dx\!\!\!&=&\!\!\!\frac{\bar{\delta}^q_0}{n+1}\Big[
\tilde{x}_1^{n+1}{\rm Ci}(\pi q \tilde{x}_1)
-\frac{\tilde{x}_1^{n}}{\pi q}\sin(\pi q\tilde{x}_1)
+\frac{n\bar{\delta}^{n}_0}{\pi q}
\int^{\tilde{x}_1}_0 x^{n-1}\sin(\pi q x)dx\Big]\\
&&+\frac{\delta^q_0}{n+1}\left(\gamma+\ln \pi \bar{\epsilon}+\ln\tilde{x}_1-\frac{1}{n+1}\right)\tilde{x}_1^{n+1}.
\eeas
}
The third type of integrals can be written recursively as ($k\neq 0$)
\bea\label{RecursiveCinLn}
\int_0^{\tilde{x}_1} dx\,x^n \sin(\pi k x)\ln x&=&L_{s}^{(n)}(\tilde{x}_1)+\frac{n}{\pi k}\int^{\tilde{x}_1}_0 x^{n-1}\cos(\pi k x)\ln x \\
\int^{\tilde{x}_1}_0 dx\, x^n\cos(\pi k x)\ln x&=&L_{c}^{(n)}(\tilde{x}_1)-
\frac{n}{\pi k}\int^{\tilde{x}_1}_0 x^{n-1}\sin(\pi k x)\ln x,\nonumber
\eea
with the ``rest'' functions
\beas
L_s^{(n)}(\tilde{x}_1)&=&-\frac{\tilde{x}_1^{n}}{\pi k}\Big[\ln\tilde{x}_1\cos(\pi k \tilde{x}_1)-\delta^n_0\Big({\rm Ci}(\pi k \tilde{x}_1)-\gamma-\ln \pi k\Big)\Big]\\
&&+
\bar{\delta}^n_0\Big[\frac{\tilde{x}_1^{n-1}}{(\pi k)^2}\sin(\pi k\tilde{x}_1)
-\frac{(n-1)\bar{\delta}^{n}_1}{(\pi k)^2}
\int^{\tilde{x}_1}_0 x^{n-2}\sin(\pi k x)dx\Big]\\
L_c^{(n)}(\tilde{x}_1)&=&
\frac{\tilde{x}_1^{n}}{\pi k}\Big[\ln\tilde{x}_1\sin(\pi k \tilde{x}_1)
  -\delta^n_0{\rm Si}(\pi k \tilde{x}_1)\Big]\\
&&+\bar{\delta}^n_0\Big[\frac{\tilde{x}_1^{n-1}}{(\pi k)^2}\Big(\cos(\pi k\tilde{x}_1)-\delta^n_1\Big)
-\frac{(n-1)\bar{\delta}^{n}_1}{(\pi k)^2}
\int^{\tilde{x}_1}_0 x^{n-2}\cos(\pi k x)dx\Big]. \nonumber
\eeas
Converting into sums yields
\beas
\int^{\tilde{x}_1}_0 dx\, x^n\sin_s(\pi k x)\ln x=
\bar{\delta}^k_0\sum_{j=0}^{n/2}\frac{n!}{(n-2j)!}
\frac{(-1)^j}{(\pi k)^{2j}}\left(
L_{s}^{(n-2j)}-(-1)^s\frac{n-2j}{\pi k}L_{1-s}^{(n-2j-1)}\right).
\eeas
The last integral category is handled in the next subsection as part of
an example for the calculation of matrix elements.

\subsection{Singular Matrix Elements}

For ``singular'' matrix elements one employs the
enlarged integral domain of Sec.~\ref{SecNumericalSolution} which
complicates the {\em numerator} of the integrals, i.e.~requires more
algebraic effort, but does not add to the list of integrals. Nonetheless,
evaluation is very cumbersome, as one can see from computing the generic
integral(s) Eq.~(\ref{Integral9})
$\forall\, n_1,n_2\ge 0, k_1,k_2,q\in \mathbb{Z}$ as well as $s,S\in\{0,1\}$.
We can write the inner integral recursively as 
in Eq.~(\ref{RecursiveCinLn})
where the ``rest'' functions $R_{jJ}^{(n)}(\tilde{x}_1)$ (replacing the
$L^{(n)}_j$) are
\beas
2\pi k R_{sS}^{(n)}(\tilde{x}_1)&=&-2\tilde{x}_1^n\cos(\pi k \tilde{x}_1){\rm Si}(\pi q \tilde{x}_1) \pm\sum_{\mp}\bar{\delta}^k_{\pm q}\Big\{\delta^n_0\,{\rm Si}(\pi(k\mp q)\tilde{x}_1)+\bar{\delta}^n_0\,\Gamma_{\mp}(\tilde{x}_1)\Big\}\\
\\ 
2\pi k R_{cS}^{(n)}(\tilde{x}_1)&=&2\tilde{x}_1^n\sin(\pi k \tilde{x}_1){\rm Si}(\pi q \tilde{x}_1)
-(\delta^k_q-\delta^k_{-q})\left(\frac{\bar{\delta}^n_0}{n}
+\delta^n_0\ln\tilde{x}_1\right)\tilde{x}_1^n\Big\}\\
&& \mp\sum_{\mp}\bar{\delta}^k_{\pm q}\Big\{\delta^n_0
\Big[{\rm Ci}(\pi|k\mp q|\tilde{x}_1)-\gamma-\ln(\pi|k\mp q|)\Big]
+\bar{\delta}^n_0\,\bar{\Gamma}_{\mp}(\tilde{x}_1)\Big\}\\
R_{sC}^{(n)}(\tilde{x}_1)&=&\delta^q_0\Big[\left(\gamma+\ln(\pi\bar{\epsilon})
  \right)\int_0^{\tilde{x}_1} x^n \sin(\pi k x)dx+\int_0^{\tilde{x}_1} x^n \sin(\pi k x)\ln x\, dx\Big]
\\
&&+\frac{\bar{\delta}^q_0}{2\pi k}\Bigg\{
-2\tilde{x}_1^n\cos(\pi k \tilde{x}_1){\rm Ci}(\pi q \tilde{x}_1)
+(\delta^k_q+\delta^k_{-q})\left(\frac{\bar{\delta}^n_0}{n}
+\delta^n_0\ln\tilde{x}_1\right)\tilde{x}_1^n\Bigg\}\\
&&+\sum_{\mp}\bar{\delta}^k_{\pm q}\Big\{\delta^n_0\tilde{x}_1^n
\Big[{\rm Ci}(\pi|k\mp q|\tilde{x}_1)-\gamma-\ln(\pi|k\mp q|)\Big]
+\bar{\delta}^n_0\,\bar{\Gamma}_{\mp}(\tilde{x}_1)\Big\}\\
\eeas
\beas
R_{cC}^{(n)}(\tilde{x}_1)&=&\delta^q_0\Big[\left(\gamma+\ln(\pi\bar{\epsilon})
  \right)\int_0^{\tilde{x}_1} x^n \cos(\pi k x)dx+\int_0^{\tilde{x}_1} x^n \cos(\pi k x)\ln x\, dx\Big]
\\
&&+\frac{\bar{\delta}^q_0}{2\pi k}\Bigg\{2\tilde{x}_1^n\sin(\pi k \tilde{x}_1){\rm Ci}(\pi q \tilde{x}_1) -\sum_{\mp}\bar{\delta}^k_{\pm q}\Big\{\delta^n_0{\rm Si}(\pi(k\mp q)\tilde{x}_1)-\bar{\delta}^n_0\,\Gamma_{\mp}(\tilde{x}_1)\Big\},
\eeas
where
\beas
\Gamma_{\mp}(\tilde{x}_1)&:=&\frac{\tilde{x}_1^{n-1}}{\pi(k\mp q)}\left(\cos(\pi(k\mp q)\tilde{x}_1)-\delta^{n}_1\right)-\frac{(n-1)\bar{\delta}^{n}_1}{\pi(k\mp q)}
\int^{\tilde{x}_1}_0 x^{n-2}\cos(\pi (k\mp q) x)dx,
\\
\bar{\Gamma}_{\mp}(\tilde{x}_1)&:=&\frac{\tilde{x}_1^{n-1}}{\pi(k\mp q)}\sin(\pi(k\mp q)\tilde{x}_1)-\frac{(n-1)\bar{\delta}^{n}_1}{\pi(k\mp q)}
  \int^{\tilde{x}_1}_0 x^{n-2}\sin(\pi (k\mp q) x)dx.
\eeas
The integrals in these expressions are sums of powers of $\tilde{x}_1$ and
trig functions, Eqs.~(\ref{xnCinIntegrals}) plus integrals with an additional
logarithm. In other words, integrals of categories one and three. 
Then
\beas
\int^{\tilde{x}_1}_0 \!\!\!\!dx\, x^n\sin_s(\pi k x){\rm Si}_S(\pi q x)\!\!\!&=&\!\!\!
\bar{\delta}^k_0
\sum_{j=0}^{n/2}\frac{n!}{(n-2j)!}
\frac{(-1)^j}{(\pi k)^{2j}}\left(
R_{sS}^{(n-2j)}-(-)^s\frac{n-2j}{\pi k}R_{1-s,S}^{(n-2j-1)}\right)\\
&&+\delta^k_0\delta^s_0\int_0^{\tilde{x}_1}dx\,x^n{\rm Si}_S(\pi q x).
\eeas
With the inner integral evaluated, we can now use
\[
\int^1_0 x_1^{n} \sin_s(\pi k_1 x_1)f(1-x_1)dx_1=\sum_{p=0}^n(-1)^{k_1+s+p}
\left({n\atop p}\right)\int_0^1 \tilde{x}^p_1\sin_s(\pi k_1 \tilde{x}_1)
f(\tilde{x}_1)d\tilde{x}_1
\]
and the occasional trig identity to write our integral (\ref{Integral9})
as a linear combination of the four categorized integrals above.

\subsection{Regular Matrix Elements}
\label{AppxRegular} 

By comparison, the regular parton-number conserving interaction is
easy to deal with, since the
term looks like 
\[
\frac{g^2N}{\pi(x_1+x_2)^2}\int_0^{x_1+x_2} dy\, \phi(y,x_1+x_2-y, x_3,...x_b).
\]
It is easy to see that the term vanishes for $r=2$.
The general result ($r>3$) is essentially the integral
\[
\int_0^{x_1+x_2} dy \,\sin_s\pi(n_1-n_2)y=
(-1)^s\left[\frac{1}{\pi(n_1-n_2)}\sin_{1-s}(\pi(n_1-n_2)(x_1+x_2))
  -\delta^s_0\right].
\]
The projection integral is cumbersome since we pick up
the factor $\frac{1}{(x_1+x_2)^2}$ which becomes part of the $x_1$ and $x_2$
integrations and is singular at the lower limit.
As before, we assume the spectator momenta have been separated and integrated
out, producing linear combinations of polynomials and sinusoidals
of $x_1$ and $x_2$, such that the arising types of integrals are limited to
\beas
\int_0^1 dx_1 \int^{1-x_1}_0 dx_2 \frac{\sin_s(\pi(k_1 x_1+k_2 x_2)}{(x_1+x_2)^n},
\qquad n=0,1,2; k_i\in\mathbb{Z}.
\eeas

Note that the asymptotically massless fermionic
states $|0^{r-1}\rangle$ have
a regular matrix element $\langle 0^{r-1}|\hat{H}_{PC}|0^{r-1}\rangle$ that is easily calculated due to
\[
\int_0^1 dx_1 \int_0^{1-x_1}dx_2\frac{1}{x_1+x_2}\int_0^{1-x_1-x_2}dx_3\cdots\int_0^{1-\sum_j^{r-2}x_i} dx_{r-1} = \frac{1}{(r-2)!}.
\]
The norm for these constant states is the
square-root of the inverse integration volume $r!$, so we obtain
Eq.~(\ref{ZeroMass}),
which leads, as pointed out above, to a
good estimate for the ground state masses
in the fermionic sectors.

\subsection{Parton-Number Violating Matrix Elements}
\label{AppxPNV} 

The calculation of the parton-number violating matrix elements involves
integrals similar to the regular matrix elements of Appx.~\ref{AppxRegular}.
To wit, for incoming cosines we have
\beas
&&\int_0^1 dx_1\int_0^1 dy_1\int_0^{1-y_1} dy_2\int_0^{1-y_1-y_2} dy_3\cos\pi m_1x_1
\cos\pi(n_1y_1+n_2y_2+n_3y_3)\\
&&\qquad\qquad\qquad\qquad\qquad\times\left[\frac{1}{(y_1+y_2)^2}-
  \frac{1}{(y_2+y_3)^2}\right]\delta(x_1-y_1-y_2-y_3)\\
&&=\int_0^1 dx_1\cos\pi m_1x_1 \int_0^{x_1} dy_1 \left[I_{PNV,1c}(x_1,y_1)-I_{PNV,2c}(x_1,y_1)\right],\\
\eeas
where
\bea\label{FirstTerm}
I^{(n_2 = n_3)}_{PNV,1c}(x_1,y_1)
\!\!&:=&\cos\pi(n_3x_1+(n_1-n_3)y_1)\left[\frac{1}{y_1}-\frac{1}{x_1}\right]\\
I^{(n_2\neq n_3)}_{PNV,1c}(x_1,y_1)
\!\!&:=&\frac{\cos\pi(n_3x_1\!+\!(n_1\!-\!n_3)y_1)}{y_1}-\frac{\cos\pi(n_2x_1\!+\!(n_1\!-\!n_2)y_1)}{x_1}-\pi(n_2\!-\!n_3)\nonumber\\
&&\times\left\{\cos\pi((n_2-n_1)y_1-n_3x_1)[{\rm Si}(\pi(n_2-n_3)x_1)-{\rm Si}(\pi(n_2-n_3)y_1)]\right.\nonumber\\
&&\left.\qquad\!\!\!-\sin\pi((n_2\!-\!n_1)y_1\!-\!n_3x_1)[{\rm Ci}(\pi(n_2\!-\!n_3)x_1)-{\rm Ci}(\pi(n_2\!-\!n_3)y_1)]\right\}\nonumber
\label{SecondTerm1}
\\
I^{(n_2\neq n_3)}_{PNV,2c}(x_1,y_1)
&:=&\frac{1}{\pi(n_3-n_2)}\frac{\sin\pi(n_3x_1+(n_1-n_3)y_1)}{(x_1-y_1)^2}+(n_2\leftrightarrow n_3)\\
\label{SecondTerm2}
I^{(n_2=n_3)}_{PNV,2c}(x_1,y_1)&:=&
\frac{\cos\pi(n_3x_1+(n_1-n_3)y_1)}{x_1-y_1}.
\eea
The next step is the $y_1$ integration which yields
{\small
\beas
\int_{\epsilon}^{x_1}\!\!\!I^{(n_2 = n_3,n_1 \neq n_3)}_{PNV,1c}dy_1\!\!\!
&=&\frac{1}{\pi(n_1-n_3)}\frac{\sin\pi n_3x_1-\sin \pi n_1x_1}{x_1}
-\sin\pi n_3x_1{\rm Si}(\pi(n_1-n_3)x_1)\\
&&+\cos\pi n_3x_1[{\rm Ci}(\pi(n_1-n_3)x_1)-{\rm Ci}(\pi(n_1-n_3)\epsilon)]
  \\
\int_{\epsilon}^{x_1}\!\!\!I^{(n_2 = n_3,n_1 = n_3)}_{PNV,1c}dy_1\!\!\!&=&
\cos\pi n_3x_1[\ln x_1 -\ln\epsilon -1]\\
\int_{\epsilon}^{x_1}\!\!\!I^{(n_2 \neq n_3,n_2 \neq n_1)}_{PNV,1c}dy_1\!\!\!&=&\!\!\!
\delta^{n_1}_{n_3}\cos\pi n_3x_1[\ln x_1 -\ln\epsilon]-\bar{\delta}^{n_1}_{n_3}
\Big\{\sin(\pi n_3 x_1){\rm Si}(\pi (n_1-n_3)x_1)\\
&&\qquad\qquad\qquad-\cos(\pi n_3 x_1)[{\rm Ci}(\pi (n_1-n_3)x_1)-{\rm Ci}(\pi (n_1-n_3)\epsilon)]\Big\}\\
&&\!\!\!+\frac{\sin(\pi n_1x_1)-\sin(\pi n_2x_1)}{\pi(n_2-n_1)x_1}\\
&&\!\!\!-\pi(n_2\!-\!n_3)\Bigg[
    \frac{{\rm Si}(\pi(n_2-n_3)x_1)}{\pi(n_2-n_1)}[\sin(\pi (n_2-n_1-n_3)x_1)
      -\sin(\pi n_3x_1)]\\
    &&\qquad\qquad\qquad\qquad-I_{cS}(x_1,n_2-n_1,n_2-n_3,-n_3x_1)\\
    &&\qquad\qquad\,\,\,+\frac{{\rm Ci}(\pi(n_2-n_3)x_1)}{\pi(n_2-n_1)}[\cos(\pi (n_2\!-\!n_1\!-\!n_3)x_1)-\cos(\pi n_3x_1)]\\
    &&\qquad\qquad\qquad\qquad+I_{sC}(x_1,n_2-n_1,n_2-n_3,-n_3x_1)\Bigg]\\
\int_{\epsilon}^{x_1}\!\!\!I^{(n_2 \neq n_3,n_2 = n_1)}_{PNV,1c}dy_1&=&
\cos(\pi n_3 x_1)[{\rm Ci}(\pi (n_1-n_3)x_1)-{\rm Ci}(\pi (n_1-n_3)\epsilon)]\\
&&-\sin(\pi n_3 x_1){\rm Si}(\pi (n_1-n_3)x_1)
-\cos(\pi n_2x_1)\\
&&+\cos(\pi n_3x_1)[\cos(\pi (n_2\!-\!n_3)x_1)\!-\!1]
-\sin(\pi n_3x_1)\sin(\pi (n_2\!-\!n_3)x_1),
\eeas
}
where
\[
I_{sS}(k,q,b)=\int_0^x \sin_s(\pi(kx'+b)){\rm Si}_S(\pi q x') dx'.
\]
The contributions from the second terms,
Eqs.~(\ref{SecondTerm1})\&(\ref{SecondTerm2}),
are similar, but only contribute when $n_1\neq n_3$.
Incidentally, a multi-dimensional
{\em numerical} integration might have trouble converging, since one integrates
divergent functions in intermediate steps.
For incoming sine wavefunctions we obtain similar results.

From the expressions it is clear that the final integration over $x_1$ can be
expressed in terms of definite integrals of the four types discussed above.
The simplest are\footnote{We have to be careful at the lower limit here, and use $\lim_{\epsilon\rightarrow 0}\int_{\epsilon}^1$ whence, e.g.
\beas
\lim_{\epsilon\rightarrow 0} {\rm Ci}(\pi(k-q)\epsilon)+{\rm Ci}(\pi(k+q)\epsilon)-2{\rm Ci}(\pi q\epsilon)\cos(\pi k \epsilon)
&=&-\ln \left|\frac{k^2}{q^2}-1\right|.  
\eeas
Dropping the $\epsilon^2$
term, the cosine doesn't change sign with $k$,
unlike its counterpart at the upper limit. 
Note that we are assuming $k,q\in\mathbb{Z}$ and that expressions for identical arguments have to be worked out separately.
}
\beas
D_{cC}(k,q)\!:=\!\int_0^1 \cos(\pi k x){\rm Ci}(\pi q x) dx \!\!&=&\!\!\frac{\bar{\delta}^k_0}{2\pi k} \left[{\rm Si}(\pi(k-q))+{\rm Si}(\pi(k+q))
  \right]+\delta^k_0{\rm Ci}(\pi q)\\
D_{sS}(k,q)\!:=\!\int_0^1 \sin(\pi k x){\rm Si}(\pi q x) dx \!\!&=&\!\! \frac{\bar{\delta}_0^{k}}{2\pi k} \left[{\rm Si}(\pi(k\!-\!q))-{\rm Si}(\pi(k\!+\!q))+2(-)^k{\rm Si}(\pi q)\right]\\
D_{cS}(k,q)\!:=\!\int_0^1 \cos(\pi k x){\rm Si}(\pi q x) dx \!\!&=&\!\!
\frac{\bar{\delta}^k_0}{2\pi k}\Big[{\rm Ci}(\pi|k-q|)-{\rm Ci}(\pi|k+q|)
  \\
  &&\left.-\ln\left|\frac{k-q}{k+q}\right|\right]+\delta^k_0
\left[{\rm Si}(\pi q)+\frac{(-1)^q-1}{\pi q}\right]\\
D_{sC}(k,q)\!:=\!\int_0^1 \sin(\pi k x){\rm Ci}(\pi q x) dx \!\!&=&\!\!\frac{1-\delta^k_0}{2\pi k} \Big[{\rm Ci}(\pi|k-q|)+{\rm Ci}(\pi|k+q|)\\
  &&\left.-2{\rm Ci}(\pi |q|)(-1)^k-\ln\left|\frac{k^2}{q^2}-1\right|\right].
\eeas
Note that $D_{cC}(0,q\rightarrow 0)=\gamma+\ln(\pi q)-1$ is divergent, whereas
\[
D_{sC}(k,0)=\frac{1}{\pi k}\left[{\rm Ci}(\pi |k|)-\ln |k|-(-1)^k(\gamma+\ln\pi)
  -\ln q((-1)^k-1) \right]
\]
is finite only for even $k$.

\subsection{Mass Term Elements}

The mass term elements are dramatically simplified by using the union of all
unique Hilbert spaces, cHS, as the integration domain,
see Sec.~\ref{SecNumericalSolution}. They read
\[
\langle \vec{\mu}|\hat{H}_m|\vec{\nu}\rangle=\frac{m^2}{2}\int_0^1 \frac{dx_1}{x_1}\int_0^{1-x_1} dx_2
\cdots\int_0^{1-\sum_i^{r-1} x_i} dx_{r-1} \sin_I(\pi \vec{\mu}\cdot \vec{x})
  \sin_I(\pi \vec{\nu}\cdot \vec{x}), 
\]
where $I$ is the ${\cal I}$ quantum number of the sector, $\sin_{-1}=\sin$,
and $\sin_{+1}=\cos$.    

\end{appendix}

\end{document}